\documentclass[aps,pre,twocolumn,superscriptaddress,showpacs,showkeys,a4paper]{revtex4}
\usepackage[english]{babel}
\usepackage{amsmath}
\usepackage{float}
\usepackage{graphicx}
\usepackage{grffile}
\usepackage{hyperref}
\usepackage{url}
\usepackage{color}

\newcommand{\DS}[1]{\texttt{#1}}
\newcommand{\CDA}[1]{\texttt{#1}}

\begin{document}

\title{Community detection in networks: Structural communities versus
  ground truth}

\author{Darko Hric}
\affiliation{Department of Biomedical Engineering and Computational
  Science, Aalto University School of Science, P.O.  Box 12200,
  FI-00076, Finland}
\author{Richard K. Darst}
\affiliation{Department of Biomedical Engineering and Computational
  Science, Aalto University School of Science, P.O.  Box 12200,
  FI-00076, Finland}
\author{Santo Fortunato}
\affiliation{Department of Biomedical Engineering and Computational
  Science, Aalto University School of Science, P.O.  Box 12200,
  FI-00076, Finland}

\begin{abstract}
Algorithms to find communities in networks rely just
on structural information and search for cohesive
subsets of nodes. On the other hand, most scholars implicitly or
explicitly assume that structural communities represent groups of nodes with
similar (non-topological) properties or functions. 
This hypothesis could not be
verified, so far, because of
the lack of network datasets with information on the
classification of the nodes. We show that traditional community
detection methods fail to find the metadata groups in many large
networks. Our results show that there is a marked separation between
structural communities and metadata groups, in line with recent findings. That means that either
our current modeling of community structure has to be
substantially modified, or that metadata groups may not be
recoverable from topology alone. 
\end{abstract}

\pacs{89.75.Hc}
\keywords{Networks, community structure}
\maketitle

\section{Introduction}

Detecting communities in networks is one of the most popular topics of
network science~\cite{fortunato10}. Communities are
usually conceived as subgraphs of a network, with a high density of
links within the subgraphs and a comparatively lower density between
them. The existence of community structure indicates that the
nodes of the network are not homogeneous but divided into
classes, with a higher probability of connections between
nodes of the same class than between nodes of different classes. This
can have various reasons. In a social network, for instance, the
communities could be groups of people with common interests, or
acquaintanceships; in protein interaction networks they might indicate
functional modules, where proteins with the same function 
frequently interact in the cell, hence they share more links; in the
web graph, they might be web pages dealing with similar topics,
which therefore refer to each other.

One of the drivers of community detection is the possibility to
identify node classes, and to infer their attributes, when they are
not directly accessible via experiments or other channels. However,
community detection algorithms are usually informed only by the network
structure (in many cases this is all the information available). So, one
postulates that structural communities coincide or are strongly
correlated with the node classes, which correspond to their intrinsic
features or functions. In a sense, the field has been silently
assuming that structural communities {\it reveal} the non-topological classes.
This is confirmed by the fact that community detection algorithms are typically
tested on a (low) number of real networks where the classification of
the nodes is available, such as, e.g.\ Zachary's karate club~\cite{zachary77}, Lusseau's
dolphins network~\cite{lusseau03} and the college football network~\cite{girvan02}.
This way, one implicitly tunes hypotheses and/or parameters such to
get the best match between the communities detected by the method and
the metadata groups of those systems.

Our goal is testing this basic hypothesis. This has finally become possible, due to
the availability of several large datasets with 
information on the classification of the nodes (the node \textit{metadata}). 
In recent work, Yang and Leskovec have studied the topological
properties of metadata groups in social, information and
technological networks~\cite{yang12,yang13,yang14}. They found that
they have peculiar properties, some of which are in contrast with the
common picture of community structure. For instance, it seems that
overlapping communities have a higher density of links in the
overlapping than in the non-overlapping parts~\cite{yang14}, which is
the opposite of what one usually thinks. 

In this paper we will compare the community structure detected by popular community detection
algorithms on a collection of network datasets with the metadata node groups
of the networks. Comparisons will be carried out both at the level of
the whole partition, and at the level of the individual
communities. We find that the match between topological and supposed
``ground truth communities'' (metadata groups) is not good, for all methods employed in the
analysis. This questions the usefulness of (purely topological) community detection algorithms to
extrapolate the hidden (non-topological) features of the nodes.

Before we proceed, it is worthwhile to clarify some nomenclature.  We
take \textit{community} to represent a connected subgraph
with a density of internal links which is appreciably higher than the
density of external links. The term \textit{cluster} is often used
interchangeably with ``community'' within the physics literature, but
has a more general meaning within computer science.  For clarity, the
sets of nodes derived from the network metadata (which are hopefully
detected by methods) are known as \textit{metadata groups}. These are
not assumed to represent structural communities until proven. The
term \textit{ground truth} is used in other literature to refer to
these metadata groups in order to invoke the concept of a true result
to which we will attempt to match.  We avoid the term here because of
the reason above.  The term \textit{partition} formally refers to a
complete, non-overlapping set of communities, but in this work we
loosen the definition to any set of communities.  While some datasets
do have strict partitions, others can have overlapping nodes (nodes in
multiple groups) and nodes in no groups.

In Section~\ref{sec:data} we will introduce our collection of datasets and the
community detection methods used in the
study. Section~\ref{sec:properties} reports some basic structural
properties of the metadata groups. Sections \ref{sec:part-analysis} and \ref{sec:jaccard}
expose the results of the comparison between detected communities and
metadata groups, both at the level of the partition as a
whole (Section \ref{sec:part-analysis}) and at the level of the individual groups
(Section \ref{sec:jaccard}). In Section \ref{sec:conclusions} we will discuss the implications of the
results.

\section{Data and community detection methods}
\label{sec:data}

\subsection{Network datasets}

We collected many networks with node metadata that can be used for
creating different node classes to approximate communities, which we
refer to as \textit{metadata groups}.  These datasets can roughly be classified in two
groups: classical and big datasets. Full details on all datasets can
be found in Appendix~\ref{sec:appendix-datasets}.

The first group contains real and synthetic networks that have
regularly been used for testing community
detection algorithms. Zachary's karate club
network (\texttt{karate}) is a classic
testbed for community detection algorithms~\cite{zachary77}: it has two natural
communities, corresponding to the split of the club in two factions. 
So is \texttt{football}, which represents matches played between US college
football teams in year 2000~\cite{girvan02}; the metadata groups are team conferences. \texttt{polblogs} is the
network of political blogs after the 2004 elections in the US~\cite{adamic05}, grouped by
political alignment. \texttt{polbooks} represent copurchased books on
politics on Amazon bookstore around the time of 2004 elections, and grouped by
political alignment~\cite{krebs06}. We also used a state-of-the-art artificial
network with built-in (topological) communities,
the LFR benchmark~\cite{lancichinetti08}, with 1000 vertices, small communities, and
mixing parameter of $\mu=0.5$ (\texttt{lfr}).

The second group contains more recent and challenging networks. The Debian package
dependencies (\texttt{dpd}) are dependencies of software packages in Debian
Linux distribution, grouped by crowd-sourced tag assignment. The Pretty Good Privacy
network (\texttt{pgp}) contains email addresses with signatures
between them, with groups represented by the email
domains~\cite{garfinkel95}. The Internet topology at the level of autonomous systems
(\texttt{as-caida}) is
collected by the CAIDA project, and is grouped by countries~\cite{caida-as-rel,caida-ip-routed}. The Amazon product
copurchasing network (\texttt{amazon}) has groups of product
categories~\cite{amazon}. \texttt{anobii} is
a book recommendation social network popular in Italy, where users can join
groups~\cite{aiello12,aiello10}.  The \texttt{dblp} network of coauthorships in computer
science literature has publication venues as groups~\cite{backstrom06}. The Facebook
university networks (\texttt{fb100}) consist of 100 separate networks
of Facebook users at US universities from 2005~\cite{traud12}. The multiplicity was
used to provide statistics.  The groups are freely entered by users and are formed with different
criteria (such as field of study or graduation year) provided in the
node metadata.
The network of Flickr users (\texttt{flickr}) consists of photo-sharing users who
join user groups to share content~\cite{mislove07}. The LiveJournal network consists of
users friendships and explicit
group memberships.  We have two independent sources for this network,
\texttt{jl-backstrom}~\cite{backstrom06} and \texttt{lj-mislove}~\cite{mislove07}, which are analyzed
separately. The Orkut social network (\texttt{orkut}) consists of users and groups
they join.

\begin{table*}
\centering
\begin{tabular}{r|rrrl}
  Name & No. Nodes & No. Edges & No. Groups & Description of group nature \\
  \hline
  lfr & 1000 & 9839 & 40 & artificial network (lfr, 1000S, $\mu=0.5$) \\
  karate & 34 & 78 & 2 & membership after the split \\
  football & 115 & 615 & 12 & team scheduling groups \\
  polbooks & 105 & 441 & 2 & political alignment \\
  polblogs & 1222 & 16782 & 3 & political alignment \\
  dpd & 35029 & 161313 & 580 & software package categories \\
  as-caida & 46676 & 262953 & 225 & countries \\
  fb100 & 762--41536 & 16651--1465654 & 2--2597 & common
students' traits \\
  pgp & 81036 & 190143 & 17824 & email domains \\
  anobii & 136547 & 892377 & 25992 & declared group membership \\
  dblp & 317080 & 1049866 & 13472 & publication venues \\
  amazon & 366997 & 1231439 & 14--29432 & product categories \\
  flickr & 1715255 & 22613981 & 101192 & declared group membership \\
  orkut & 3072441 & 117185083 & 8730807 & declared group membership \\
  lj-backstrom & 4843953 & 43362750 & 292222 & declared group membership \\
  lj-mislove & 5189809 & 49151786 & 2183754 & declared group membership \\
\end{tabular}
\caption{Basic properties of all datasets used in this analysis.
  \texttt{fb100} consists of 100 unique networks of
  universities, so we show the ranges of the number of nodes and edges
  of the networks, as well as of the metadata groups of the various
  partitions. \texttt{amazon} consists of a hierarchical set of 11
  group levels, we report the range of the number of groups. The number of groups is calculated after our indicated
  preprocessing (see text).}
\label{tab:dataset_properties}
\end{table*}

We present the list of datasets in Table~\ref{tab:dataset_properties}.
We converted all networks to undirected, unweighted networks, and take
their largest connected component.  This is
the largest weakly connected component (LWCC) of the directed graphs.  Any graph members outside of this
LWCC are dropped. The numbers of Table I refer to the LWCC 
of each network.

In general, the metadata groups in the data can be
disconnected within the graph.  We applied the following preprocessing steps.
Each group's connected components over the network were taken as
separate groups for the analysis.  That means that 
several distinct groups may end up having the same node
membership. On the other hand, community detection methods would not be able to
associate disconnected groups, so it is necessary to proceed like
this. Any group with less than three
members is dropped, from both the metadata partition and the
detected partition. The comparison is limited to the set of nodes
belonging to both the metadata and the detected partition after the
above preprocessing steps. 
Since in some cases the fraction of nodes
of the system belonging to such intersection can be quite low, we report results only
when it exceeds $10\%$. 

\subsection{Community detection methods}
We have a collection of community detection
methods with available codes. These methods come from a variety of
different theoretical frameworks. Some of them are designed to detect
overlapping communities, others can only deliver disjoint communities. Not all methods run to completion on the
largest datasets in a reasonable time, such dataset and method
combinations are excluded from the analysis.

\texttt{Louvain} is a greedy agglomerative method based on
modularity~\cite{blondel08}.  \texttt{Infomap}~\cite{rosvall11} is based on information compression of random
walks.  We also used a variant \texttt{InfomapSingle}~\cite{rosvall08}, which returns a
single partition instead of a hierarchy.
\texttt{LinkCommunities}~\cite{ahn10} is a method that clusters edges instead of
nodes. \texttt{CliquePerc}~\cite{palla05,reid12} scans for the regions spanned by a
rolling clique of
certain size.
\texttt{Conclude}~\cite{demeo14} uses edge centrality distances to grow communities.
\texttt{COPRA}~\cite{gregory10} uses propagation of information to classify communities
(label propagation). \texttt{Demon}~\cite{coscia12} exploits node-local
neighborhoods.  \texttt{Ganxis}~\cite{xie12} (formerly SLPA) is based on label
propagation. \texttt{GreedyCliqueExp}~\cite{lee10} begins
with small cliques as seeds and expands them optimizing a local
fitness function.

\section{Structural properties of node groups from metadata}
\label{sec:properties}

\begin{figure}
  \centerline{ \includegraphics[width=\columnwidth]{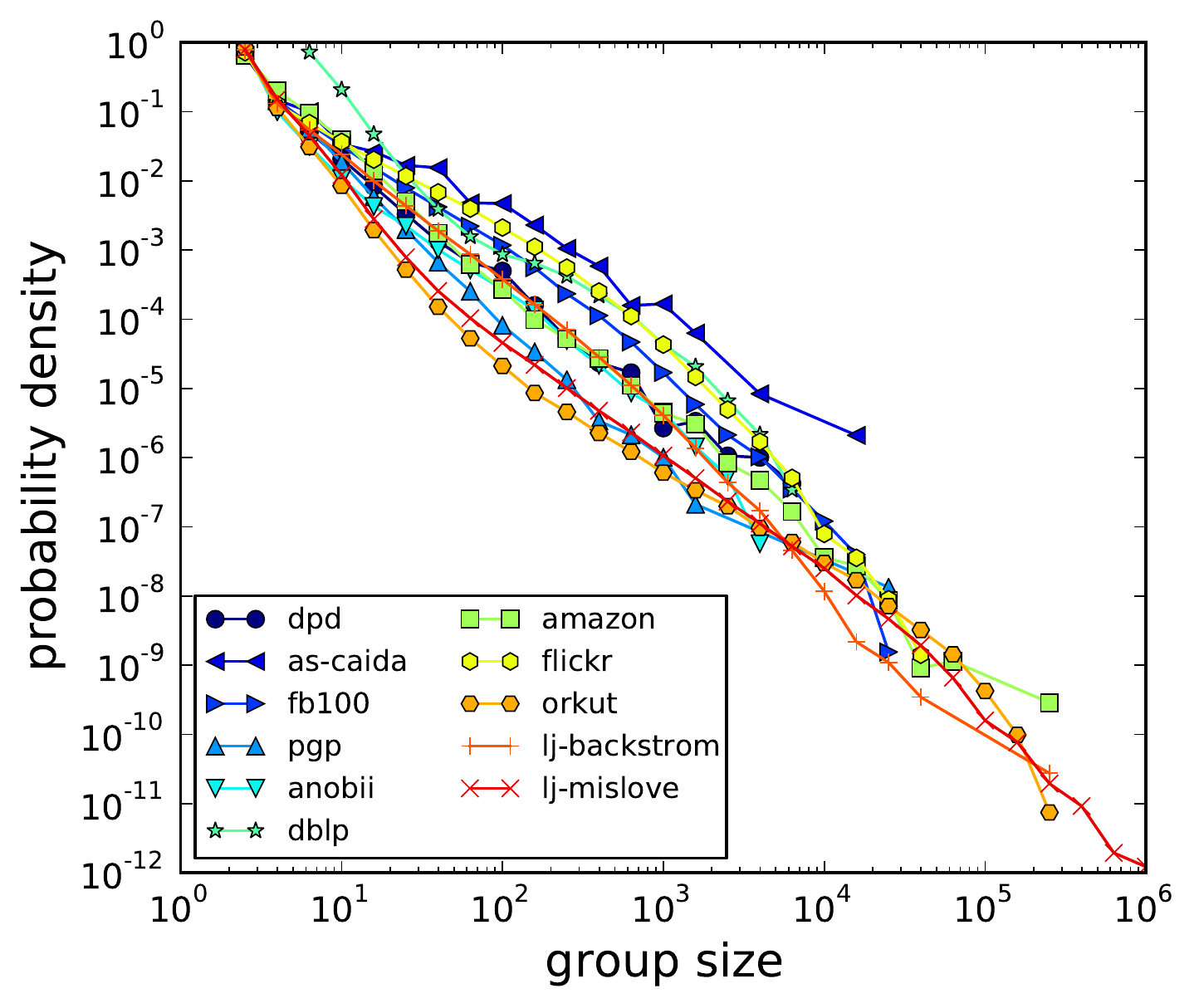} }
  \caption{(Color online) Distribution of sizes of metadata groups.
    Each curve corresponds to a specific dataset of our collection.}
  \label{fig1}
\end{figure}

Here we show some basic
topological features of the metadata groups of our
datasets. Fig.~\ref{fig1} reports the distribution of the group
sizes, which is skewed for all datasets. Power law fits of the tails
deliver exponents around $-2$. This is in agreement with the behavior
of the size distributions for the communities found by 
community detection algorithms on real networks~\cite{fortunato10}.

\begin{figure}
  \centerline{ \includegraphics[width=\columnwidth]{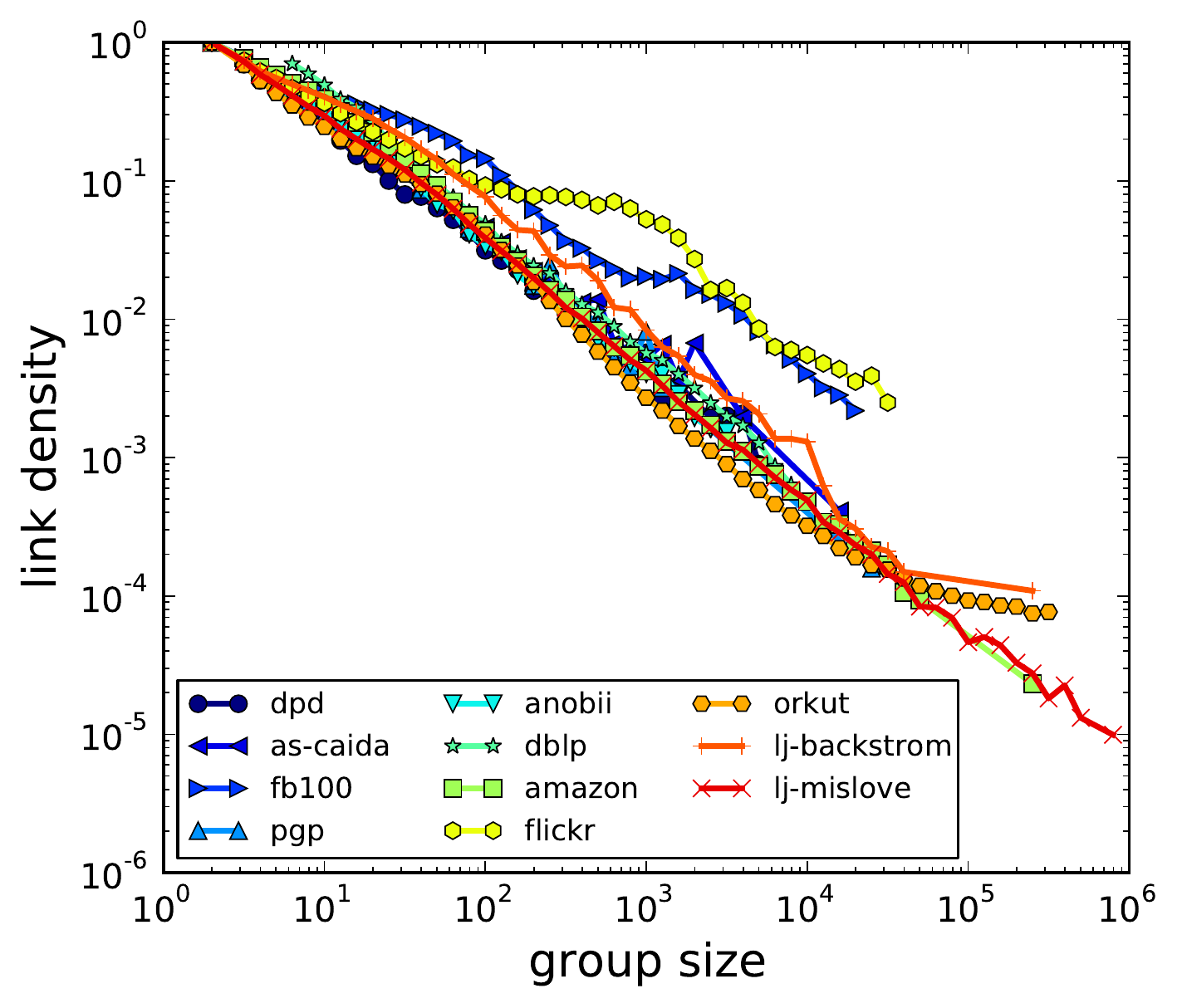} }
  \caption{(Color online) Link density 
    versus size of the metadata groups. Each curve corresponds to a specific dataset of our collection.}
  \label{fig2}
\end{figure}

The link density of a subgraph ${\cal S}$ is the ratio between the number
of links joining pairs of nodes of ${\cal S}$ and the total maximum
number of links that could be there, which is given by $n_{\cal
  S}(n_{\cal S}-1)/2$, $n_{\cal S}$ being the number of nodes of
${\cal S}$. In Fig.~\ref{fig2} we see the link density of the
metadata groups versus their sizes. Clearly, the larger the size of
the group, the lower the link density. This is because real graphs
are typically sparse, so the total number of links scales linearly
with the number of nodes. This holds for parts of the network too,
modulo small variations, so the link density decreases approximately
as a power of the number of links of the group (with exponent close
to $-1$). Since the latter is proportional to the group size, we
obtain that the link density decreases as the inverse of the group
size, as we see in Fig.~\ref{fig2}.

\begin{figure}
  \centerline{ \includegraphics[width=\columnwidth]{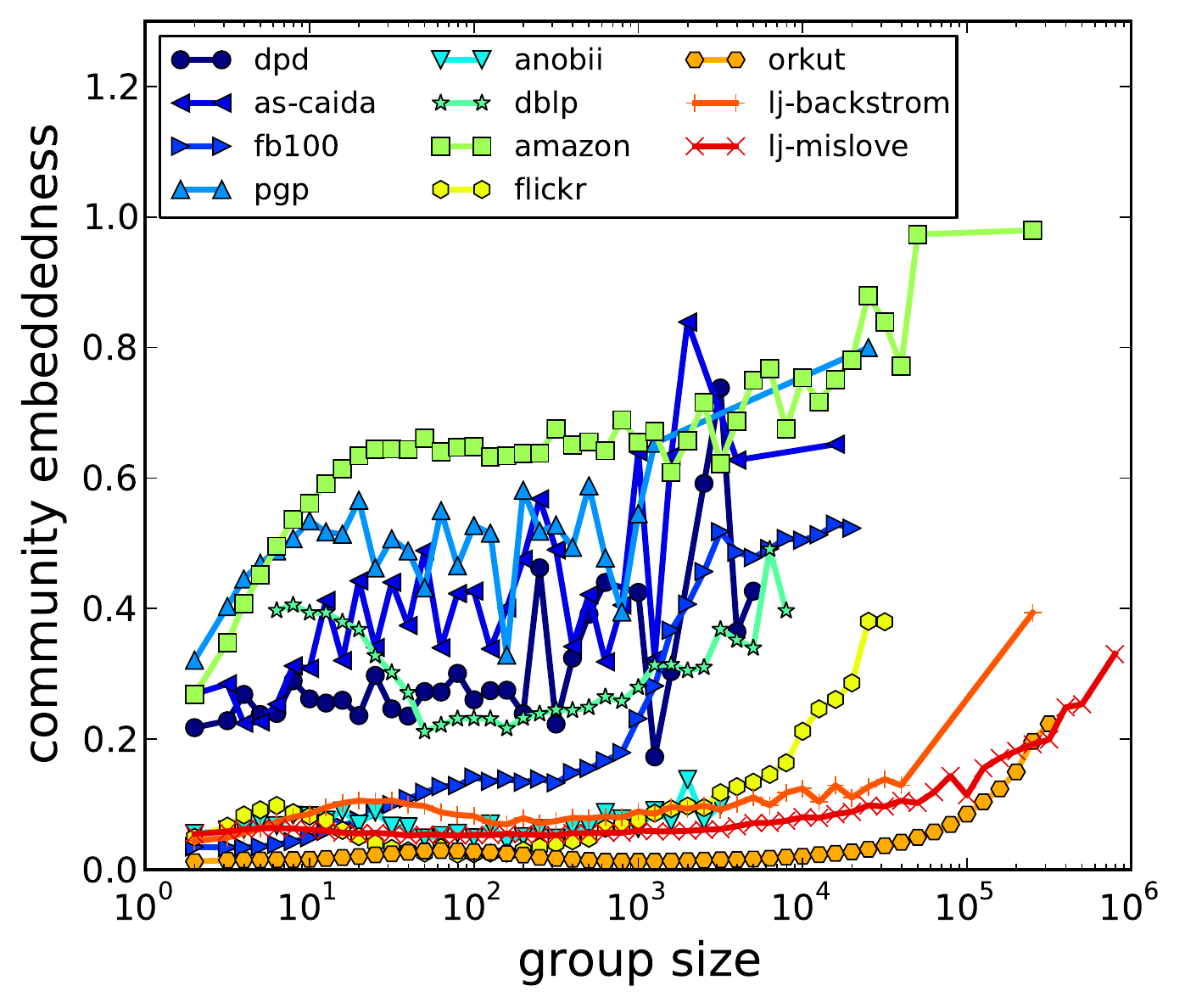} }
  \caption{(Color online) Embeddedness 
    versus size of the metadata groups. Each curve corresponds to a specific dataset of our collection.}
  \label{fig3}
\end{figure}

Finally, in Fig.~\ref{fig3} we report the relation between the
group embeddedness and its size. The embeddedness of a group
is the ratio between the internal degree of the group and the
total degree. The internal degree of a group is given by the sum of
the internal degrees of the group's nodes, i.e.\ twice the number of links
inside the group. The total degree of
the group is the sum of the degrees of its nodes. A group is
``good'' if it has high embeddedness, i.e.\ if it is well separated from
(loosely connected to) 
the rest of the graph. We notice that some of the datasets of our
collection have groups with fairly large values of the embeddedness
(e.g.\ \texttt{Amazon}), so they are fairly well separated from the other
groups. For the largest datasets we have, the online social
networks, embeddedness is very low (and fairly independent of group
size). In this case, their detection by means of community detection algorithms
is more difficult.

\section{Partition level analysis}
\label{sec:part-analysis}

The similarity of partitions can be computed in various ways (see
Ref.~\cite{fortunato10}). Here we stick to the Normalized Mutual
Information (NMI), a measure taken from information
theory~\cite{danon05}. Since the non-topological group structure of several datasets
is made of overlapping groups,
we use the generalization of the NMI proposed
by Lancichinetti \textit{et al.}, that allows for the comparison of covers
(i.e.\ of partitions into overlapping groups)~\cite{lancichinetti09}.

Many metadata as well as detected partitions do not cover all nodes
present in the network. Often these coverages mismatch, leaving many nodes
present only in one of the compared partitions. In order to circumvent
this problem we decided to follow the best possible scenario (which
generally increases the score), by using
only the nodes present in both partitions. In some cases, the fraction of
overlapping nodes was very small, so we did not calculate NMI scores if the
coverage was less than $10\%$. This only applies to comparisons
between metadata and detected partitions, for comparisons between
partitions detected with different methods we used the full sets returned
by the algorithms.

The overview of all the NMI scores is conveniently
presented in what we call ``NMI grids'', like the one in
Fig.~\ref{fig:nmi_grid_pgp}. Each grid refers to a specific
network. In addition to the NMI scores between the metadata
groups structure and the one
detected by each algorithm, we also show the similarity between
structural partitions detected by different methods. Since some methods may
deliver different hierarchical partitions, the tiles involving those
methods are further subdivided. 

\subsection{PGP NMI grid analysis}\label{section:pgp}

As an example, we provide a detailed discussion of the \texttt{pgp} NMI
grid of Fig.~\ref{fig:nmi_grid_pgp} (the others are shown in the Appendix). The
main conclusions are
consistent across all datasets, though.
Hierarchical layers were ordered by their granularity, $0$ being the lowest,
most granular one. For some algorithms layers are partitions obtained using
different parameter values
(see Appendix~\ref{sec:appendix-methods}).

First, we compare partitions returned by
different algorithms, including all returned layers (all tiles except bottom
row). On the diagonal
we have the mutual comparison of different layers delivered by the same algorithm. 
The diagonal of each tile is, of course, black, as one is comparing each
layer with itself, which yields an NMI score of 1. Off-diagonal elements show 
similarity between different layers. Most algorithms return a group of
layers which are quite similar to each other (\texttt{Infomap, Louvain,
Oslom}). Comparing the results of one algorithm versus those of other algorithms,
we can see, for instance, that the highest layer of \texttt{Infomap} is similar to some extent only to middle
layers of
\texttt{Louvain}. The lowest layer
of \texttt{CliquePerc} is much more similar to layers found by other algorithms than
to higher layers of \texttt{CliquePerc}. Layers of \texttt{LinkCommunities} (threshold values $0.25$,
$0.5$, $0.75$) show
varying behavior: the threshold value of $0.25$ yields the most
similar partitions to the ones obtained by the other algorithms, except for \texttt{Copra} and \texttt{Oslom}, and to some
extent \texttt{CliquePerc}. Lower levels of \texttt{Infomap, Louvain} and
\texttt{Oslom} tend to be more
similar to the layers returned by other algorithms. We can also draw
conclusions about the general behavior of algorithms. For instance, \texttt{Demon}
returns a partition that is not very similar to partitions returned by other
algorithms (this is more pronounced in other datasets).

The bottom row is metadata versus detected partitions, where we can see how similar
is the metadata partition to the detected ones, which is the focus
of our work. The intersections of metadata
partitions and higher order \texttt{CliquePerc} layers cover less than $10\%$ of
total nodes, so we discarded these results, indicated with hatched green in the figure.
Most of the algorithms return scores that are similar, around $0.3$. Ganxis
layers have almost the same scores (they are also very similar among
themselves) whereas \texttt{Infomap} and \texttt{Louvain} layers are very different, the lower ones
scoring better.

\begin{figure}
  \centerline{ \includegraphics[width=\columnwidth]{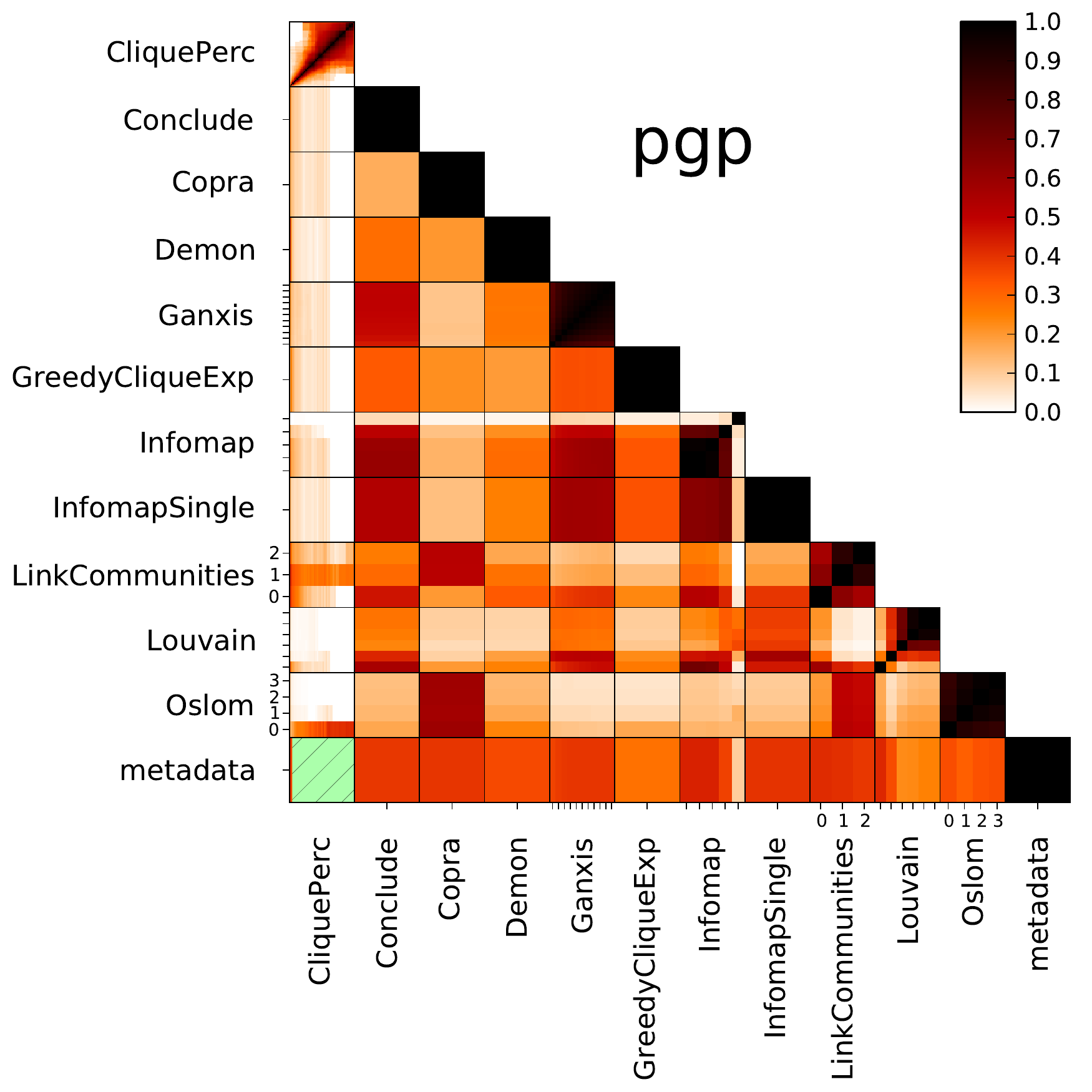} }
  \caption{(Color online) NMI grid of the \texttt{pgp}
    dataset. Each tile represents the NMI scores of the comparison of
    the structural partitions obtained from different algorithms
    and the metadata partition(s) (bottom stripe), and of the
    comparisons between partitions obtained by different algorithms.
    Each tile contains a grid within, corresponding to different
    partitions delivered by the algorithm (hierarchical levels or
    partitions obtained for given parameter choices). The color of each element of a tile
indicates the NMI score, with values discarded due to low coverage marked with
hatched green.}
  \label{fig:nmi_grid_pgp}
\end{figure}

\subsection{Overall NMI scores}

\begin{figure*}[htb]
  \centerline{
  \begin{tabular}{cc}
    \includegraphics[width=0.8\textwidth]{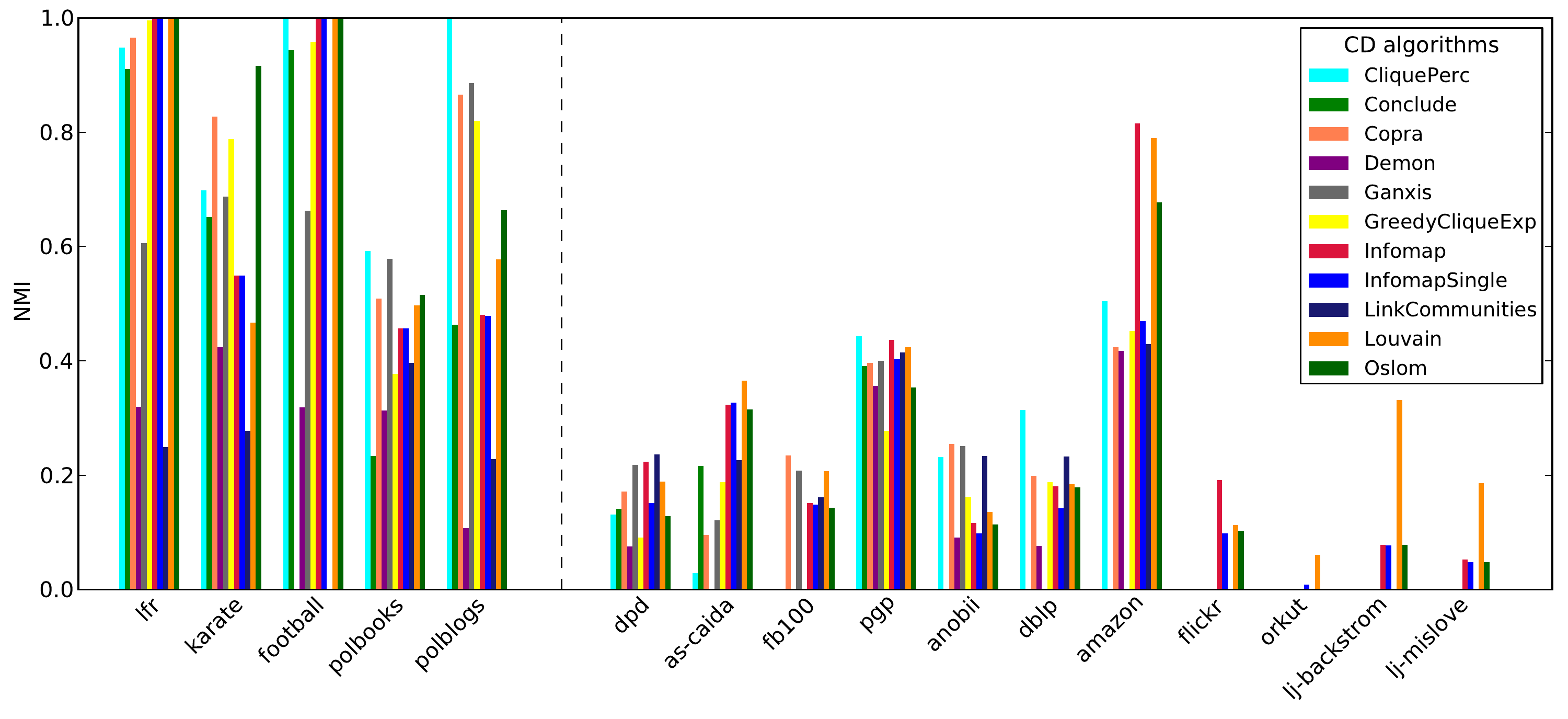}
    \end{tabular}}
  \caption{(Color online) NMI scores between structural communities and metadata
      groups for different
      networks. Scores are grouped by datasets on the \textit{x}-axis. The
      height of each column
      is the maximal NMI score between any partition layer of
      the metadata partitions and any layer returned by the community
      detection method, considering only those comparisons where the overlap of the partitions is larger than
      $10\%$ of total number of nodes.
  }
    \label{fig:bar_plot}
\end{figure*}

In order to compare how well different algorithms detect metadata groups, 
we took the best scores of each dataset-algorithm pair and present
them on Fig.~\ref{fig:bar_plot}. In real world applications one would not
know what the returned layers represent, and consequently which one of them
corresponds more truthfully to the partition one would like to
detect. So the NMI scores we derive are in general higher than
the ones obtained by comparing individual levels with each other.

The results can be separated into three groups. The highest recall of metadata groups
is in the case of the artificial dataset \texttt{lfr}, as it is
expected, since many community detection algorithms are tested on the
LFR benchmark.
The second group consists of small, classical datasets
(\texttt{karate, football, polblogs, polbooks}) that are often used for
testing community detection methods.  These NMI scores are fairly
high, but not as high as for \texttt{lfr}. The third group includes
the big datasets of our collection. Here, algorithms were
not very successful in finding the metadata groups. The only exception is
\texttt{amazon}, for which we find a much higher score than for the
others, because it has many levels for the metadata groups, some
of which turn out to be partially recoverable. Scores for the other
networks rarely go above $0.3$, for some datasets they lie even below $0.1$.

A possible explanation of the result could be that the optimization
process at the basis of several techniques is not successful, and that
the partition delivered by those methods corresponds to a value of the
measure far from the sought extreme. We reject this hypothesis
though. For one thing some
of the community detection techniques we adopted are not based on optimization
procedures (e.g. \texttt{CliquePerc}), still they do not seem to lead
to better results. Furthermore, for
\texttt{as-caida} we have computed the value of
Newman-Girvan modularity $Q$ for the metadata partition, and the ones
obtained through the \texttt{Louvain} method, corresponding to the
hierarchical level most similar to the metadata partition and to the level
yielding the best approximation to the modularity maximum. They
are $0.3839$, $0.5064$ and $0.5176$, respectively. So the values of
$Q$ of \texttt{Louvain}'s partitions are far higher than the one
corresponding to the metadata partition. 
We could not repeat this test for the other datasets because the
metadata partitions are overlapping (they are non-overlapping only
in the case of \texttt{as-caida}), while \texttt{Louvain} computes
non-overlapping partitions. Since there is no straightforward
extension of modularity to the overlapping case it is not possible to
make a meaningful comparisons of the values.

\section{Community level analysis}
\label{sec:jaccard}

The previous section shows that global measures indicate that partitions returned by community
detection methods do not align with partitions built from metadata, but what about
specific groups?  Can we detect \textit{any} of the groups well?
Are some groups reflected in the graph structure and detectable,
but lost in the bulk noise of the graph? This is what we wish to
investigate here.

The basis of our analysis is the Jaccard score between two
groups.  Let $C_i$ represent (the set of nodes of) the known
group $i$, and $D_j$ represent (the set of nodes of) the detected
community $j$.  The Jaccard score between these two sets is
defined as
\begin{equation}
  J(C_i, D_j) = \frac{\left| C_i \cap D_j \right|}{\left| C_i \cup D_j \right|},
\end{equation}
with $\left| \cdots \right|$ set cardinality, $\cap$ set intersection,
and $\cup$ set union.  The Jaccard score ranges from one (perfect
match) to zero and roughly indicates the fraction of nodes shared between the
two sets: the match quality.

The \textit{recall score} measures how well one known
group is detected.  The recall score
of one known group $C_i$ is defined as the maximal Jaccard score between
it and every detected community $D_j$,
\begin{equation}
  R(C_i) = \max_{D_j \in \{D\}} J(C_i, D_j).
\end{equation}
It is near one if the group is well detected and low otherwise.
We can study the distribution of these scores to see how many
groups can be detected at any given quality level.
Recall measures the detection of known groups, and to measure the
significance of detected communities, we can reverse the measure to
calculate a \textit{precision score}
\begin{equation}
  P(D_j) = \max_{C_i \in \{C\}} J(D_j, C_i).
\end{equation}
The precision score tells us how well one detected community
corresponds to any known group.

\begin{figure*}
  \centerline{
  \includegraphics[width=\textwidth]{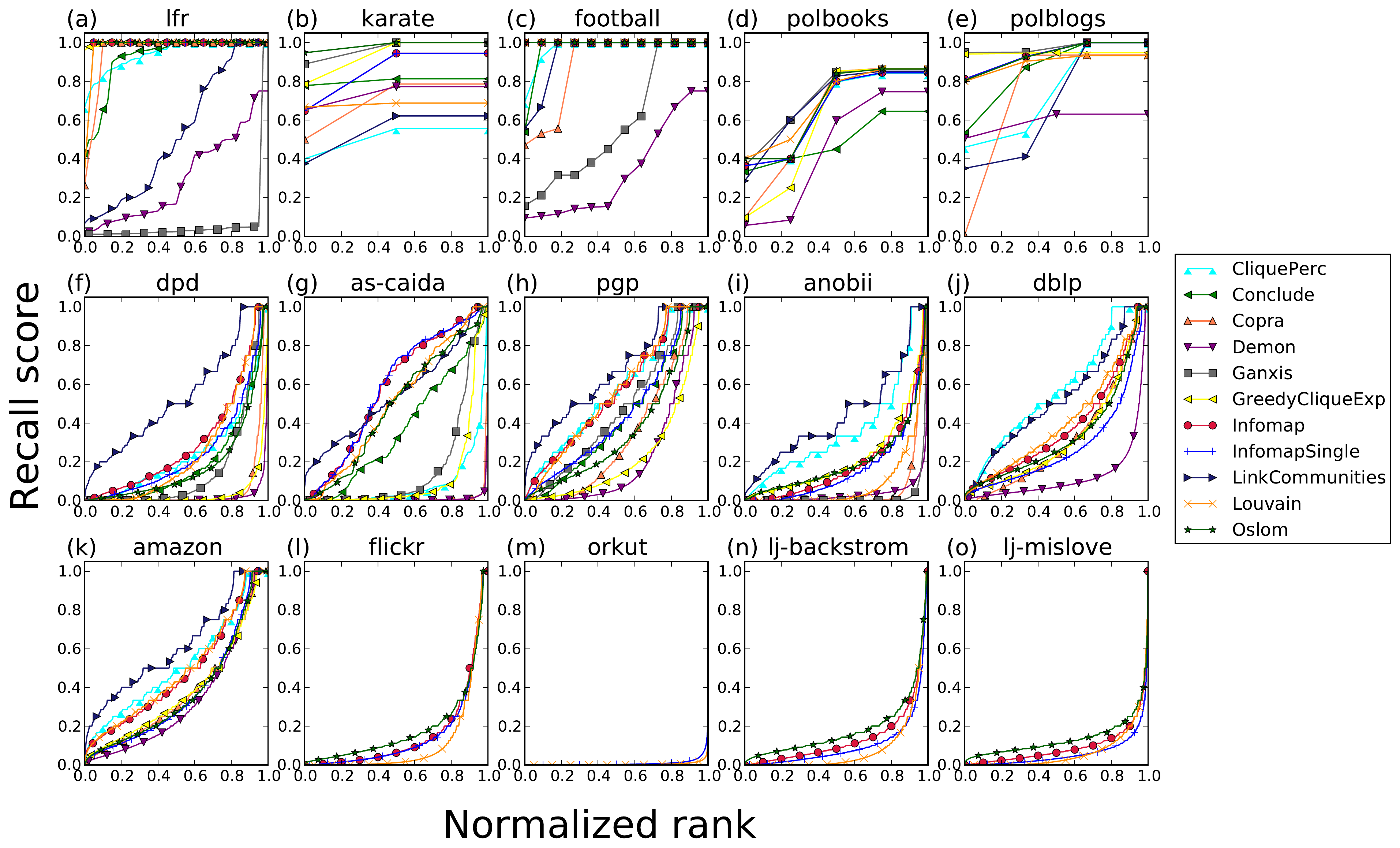}
  }
  \caption{(Color online) Recall of known groups plotted versus
    group rank (sorted by recall)
    for various datasets and methods.  Every known group is
    compared with every detected community in any layer.  We see that
    performance is usually poor (close to zero) for all networks except for
    some of the classic benchmarks (uppermost row of diagrams), 
    which are typically used to test algorithms. 
    The metadata groups of some graphs, such as
   \DS{livejournal}, \DS{orkut}, and \DS{flicker} have but a little
   overlap with the detected communities.}
  \label{fig:recl}
\end{figure*}

\begin{figure*}
  \centerline{
  \includegraphics[width=\textwidth]{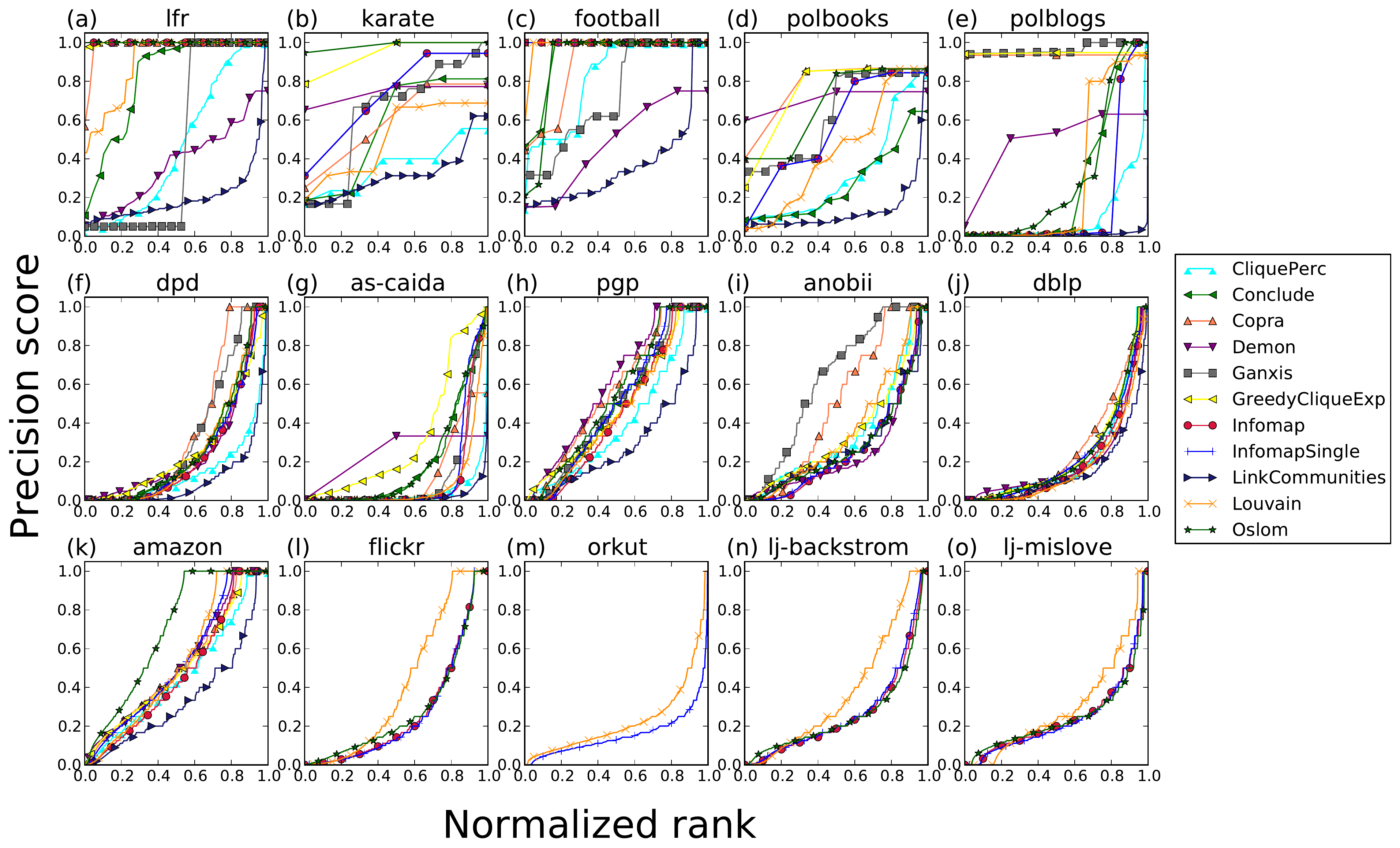}
  }
  \caption{(Color online) Precision of detected
    communities for various datasets and methods.  All detected
    communities (in any layer) are compared to all known groups.
    Results are similar as for recall (Fig.~\ref{fig:recl}).}
  \label{fig:prec}
\end{figure*}

We can now directly quantify the two conditions for good community
detection: every known group must correspond to some detected
community, and every detected community must represent some known
group.  Both of these measures are still interesting
independently: a high recall but low precision indicates that the
known groups are reflected in the network structurally, but there
are many structural communities that are not known.  We visualize the
scores by means of \textit{rank-Jaccard plots} which give an overview
of the network's detection quality.  We compute the
recall (precision) for every known (detected) group and sort
the groups in order of ascending Jaccard score.  We plot
recall (precision) vs the group rank, sorted by recall (precision) score so that the
horizontal scale is the relative group rank, i.e.\ the ratio
between the rank of the group and the number of groups (yielding a
value between $0$ and $1$).
Similar to our treatment of the partition-level analysis, we only plot
matchings whose intersection covers more than $10\%$ of total nodes in
the graph.
In our final plots, the
average value of the curve (proportional to the area under it) is the
average recall or precision score over all groups.
\def\mc#1{\multicolumn{2}{|c|}{#1}}
\begin{table*}[htb]
  \centering
  \begin{tabular}{c||cc|cc|cc|cc|cc|cc|cc|cc|cc|cc|cc|}
            & \mc{Clique P} &\mc{Conclude}&\mc{Copra}&\mc{Demon}&\mc{Ganxis}&\mc{Grd CE}&\mc{Info}&\mc{InfoS}&\mc{LinkC}&\mc{Louvain}&\mc{Oslom}\\ \hline \hline
            & R  & P  & R  & P  & R  & P  & R  & P  & R  & P  & R  & P  & R  & P  & R  & P  & R  & P  & R  & P  & R  & P  \\
lfr         &0.92&0.47&0.91&0.82&0.95&0.98&0.28&0.36&0.05&0.45&1.00&1.00&1.00&1.00&1.00&1.00&0.52&0.20&0.98&0.90&1.00&1.00\\
karate      &0.48&0.34&0.80&0.50&0.64&0.51&0.71&0.71&0.94&0.63&0.89&0.89&0.80&0.63&0.80&0.63&0.50&0.30&0.68&0.48&0.97&0.97\\
football    &0.91&0.76&0.86&0.77&0.87&0.94&0.30&0.40&0.81&0.87&0.95&0.90&0.95&0.90&0.95&0.90&0.89&0.35&0.95&0.92&0.97&0.86\\
polbooks    &0.60&0.31&0.46&0.25&0.55&0.71&0.37&0.67&0.67&0.62&0.52&0.66&0.60&0.49&0.60&0.49&0.64&0.13&0.64&0.41&0.63&0.63\\
polblogs
&0.67&0.10&0.80&0.24&0.63&0.94&0.57&0.43&0.97&0.96&0.94&0.94&0.91&0.15&0.91&0.15&0.59&0.01&0.87&0.29&0.91&0.29\\
\hline
\hline
dpd         &0.24&0.15&0.19&0.26&0.05&0.35&0.02&0.24&0.16&0.33&0.04&0.27&0.30&0.24&0.24&0.24&0.53&0.10&0.26&0.27&0.18&0.27\\
as-caida    &0.08&0.04&0.37&0.19&0.01&0.10&0.01&0.17&0.17&0.12&0.11&0.32&0.58&0.11&0.58&0.12&0.55&0.02&0.50&0.09&0.49&0.20\\
pgp         &0.58&0.41&0.43&0.50&0.34&0.57&0.23&0.61&0.47&0.51&0.24&0.49&0.57&0.46&0.45&0.51&0.67&0.31&0.57&0.47&0.37&0.55\\
anobii      &0.37&0.35&    &    &0.07&0.50&0.04&0.26&0.02&0.62&0.23&0.35&0.19&0.26&0.17&0.27&0.44&0.28&0.09&0.36&0.20&0.28\\
dblp        &0.57&0.23&    &    &0.32&0.26&0.13&0.21&    &    &0.34&0.25&0.38&0.20&0.28&0.23&0.52&0.15&0.41&0.18&0.34&0.24\\
amazon      &0.52&0.46&    &    &0.38&0.51&0.33&0.52&    &    &0.38&0.50&0.48&0.47&0.37&0.53&0.61&0.34&0.49&0.53&0.37&0.69\\
flickr      &    &    &    &    &    &    &    &    &    &    &    &    &0.16&0.28&0.15&0.27&    &    &0.12&0.42&0.19&0.29\\
orkut       &    &    &    &    &    &    &    &    &    &    &    &    &    &    &0.01&0.16&    &    &0.00&0.23&    &    \\
lj-backstrom&    &    &    &    &    &    &    &    &    &    &    &    &0.14&0.26&0.10&0.27&    &    &0.09&0.38&0.18&0.26\\
lj-mislove  &    &    &    &    &    &    &    &    &    &    &    &    &0.10&0.25&0.06&0.24&    &    &0.07&0.30&0.14&0.25
  \end{tabular}
  \caption{Average Jaccard recall (R) and precision (P) scores for all
    datasets. The scores are simple averages over all groups. 
    Horizontal lines separate the classic benchmarks from the large
    datasets.}
  \label{tab:jacc-recl}
\end{table*}
The
shape of the curve can tell us if all groups are detected equally
well (yielding a high plateau) or if there is a large inequality in detection (a high
slope).  Furthermore, this allows us to compactly represent multiple
layers.  Each independent layer of known (detected) groups can
be plotted in the same figure. 
We would generally look for the highest
curve to know if any layer has a high recall (precision).  When
computing recall (precision), unless otherwise specified, as detected
communities we consider the communities of all partitions delivered by a
method, whereas the metadata groups are those present in all
metadata partitions (if more than one partition is available in
either case). This will give us the maximum possible
recall (precision), which might be far higher than values
coming from real applications, where one typically compares
groups of the same partition (level). 

In Figs.~\ref{fig:recl} and ~\ref{fig:prec}, we show the group
recall and precision for every dataset
and every community detection method. Similar to the situation with
NMI, with the benchmark graph \DS{lfr} most methods are able to recover the true
communities. The other small graphs (b)--(e) also have most of
the structure recoverable by most methods, as they are also used as benchmarks.  However,once we get to
large data, (f)--(o), we
see a very different story.  The vast majority of these networks have
only a small number of groups detected fairly well and not many
detected communities resemble any of the metadata groups.  Many
networks, e.g.\ the online social networks, 
have almost no metadata groups reflected in the
detected communities, and vice versa, by any method.  In some networks, such as 
\DS{pgp} and \DS{amazon}, a fraction of groups are well detected.
For example, \DS{amazon} has $20\%$ of groups with a
maximal recall Jaccard score greater than $0.6$, for any method, and
is the network with
best detected communities. The performance of the methods is
comparable in most cases. \texttt{LinkCommunities} appears to give
higher recall than all other methods in most instances. However, this
is due to the fact that it usually detects many more communities than the
other methods, so there is a higher chance to find a community that
gives high overlap with the metadata groups. 
However, the precision of \texttt{LinkCommunities} 
is very low. On the largest graphs,
\texttt{Louvain} and \texttt{InfomapSingle} have consistently worse
recall than \texttt{Oslom}, but the latter has lower precision.
In Table II we report the average recall and precision for all
datasets and algorithms.

In Appendix~\ref{sec:appendix-jaccard}, we further analyze
recall and precision by narrowing the problem to group classes
selected based on size, density or
cohesiveness, or attribute types.  This includes a full analysis of
the \texttt{fb100} dataset with its specific attributes such as
student class year, field of study, or residence.  We
see that, in general, narrowing the focus to these specific classes of
groups does not allow increased predictive power on most networks.


In this section, we have broken down the community detection problem
into something more specific: instead of asking for all known
groups to match all detected communities, we are asking if (a
subset of) known groups are found by any detected communities, or
if (a subset of) detected communities correspond to real known
groups.  Even if full community detection does not have high
accuracy, a positive answer in either of these questions can produce a
result of practical use.  Instead, we see that recall and precision
are highly network-dependent, with most networks producing very low
values for both.  This even extends to social networks with
user-defined social groups.

\section{Conclusions}
\label{sec:conclusions}

Algorithms to find communities in networks are supposed to recover
groups of nodes with the same or similar features or
functions. Therefore, whenever a new algorithm is introduced, it is
usually tested not only on artificial benchmark graphs, but also on
real graphs with known node groups derived from some metadata. A good match between the
detected partition and the attribute-based partition is considered
evidence that the method is reliable. However, the correspondence
between structural communities (the ones detected by an algorithm) and
metadata groups (identified by the nodes'
attributes) has been given for granted. In this work we
have made a systematic test of this hypothesis. 

We have compared the partitions detected by several popular community detection
algorithms with the partitions resulting from non-topological features
of the nodes, on large real network datasets. We find that there is a
substantial difference between structural communities and metadata groups.
At the partition level, we find low
similarity scores. Precision and recall
diagrams show that detected communities have low overlap with the metadata groups,
and vice versa. A more detailed analysis, in which one
restricts the comparison to
groups of comparable size, link density or
embeddedness, does not reveal major improvements. Overall, results
depend more on the network than on the specific method adopted, none of
which turns out to be particularly good on any (large) dataset. 
It is fair to remark
that we have applied the community detection algorithms on the
undirected and unweighted versions of the datasets. We have done so
because few methods can handle link directions and weights, while we
wanted to test a broad class of techniques. On the other hand, it is
possible that by accounting for link directions and weights the
comparison between detected communities and metadata groups could improve.

Our results rely on the classification of the nodes,
which may not always be reliable. 
However, our collection comprises a list of very
diverse systems, and the message coming from all of them is the
same. Clearly, we cannot exclude that there may be other datasets
whose metadata groups match more closely the structural communities
found by community detection algorithms. Still, even if there were such
datasets, our point that metadata groups are not necessarily
correlated with the communities found by standard methods, contrary to
common belief, would hold. 

We remark that low similarity scores between structural and metadata
partitions were reported by Yang and Leskovec as
well~\cite{yang13}. However, that was not the focus of the work, like
in our case, and we have considered a larger set of methods and a
broader spectrum of datasets.

What kind of implications does this finding have? We envision two
possible scenarios. It may be that our conception of community
structure, which is underlying the methods currently used, is not
correct. Most algorithms usually focus on things like link densities
within the communities, or between the communities (or both). It may
be that metadata groups
are not well represented by link density, for instance, or at least
not by link density alone. Other features, like e.g.\ degree correlations,
density of loops (like e.g.\ triangles), etc. might play a
role. Indeed, Abrahao \textit{et al.}\ have shown
that structural properties of communities detected with several
algorithms are in general
different from those of metadata groups~\cite{abrahao12}. Therefore
our best bet would be carrying out a detailed investigation
of the topological properties of the metadata groups, and
trying to infer a general description from it, which could be used as
starting point of the development of new algorithms. The recent discovery of
dense overlaps between groups, for instance, might
inform new techniques, the Affiliation Graph Model
being one example of them~\cite{yang14}. 

The other possible interpretation is that metadata groups
cannot be inferred from topology alone. There certainly is a
correlation between structural and metadata groups, but it
may be not very strong. Therefore, in order to detect metadata groups,
non-topological inputs might be necessary. In the most recent
literature on community detection several such approaches have been
proposed, mostly by computer
scientists~\cite{ester06,liu09,moser09,zhou09,tang09,silva10,balasubramanyan11,atzmueller11,akoglu12,bonchi12,sun12,xu12,barbieri13,ruan13,yang13b,pool14}.

We stress, however, that structural communities are very important for
the function of a network, as they can significantly affect the
dynamics of processes taking place on the network, such as diffusion,
synchronization, opinion formation, etc. So detecting topological
communities remains crucial. We are saying that one should not expect
too much in terms of content, at least not from the algorithms
currently in use.
We hope that the scientific community of scholars working on
community detection in networks will seriously reflect on the results of our
analysis, in order to produce more reliable algorithms for applications.

\begin{acknowledgments}
  We thank Tim Evans for providing an updated version of the college
  football dataset. We acknowledge the computational resources provided by Aalto
  University Science-IT project. R.K.D.\ and S.F.\ gratefully acknowledge MULTIPLEX, grant number
317532 of the European Commission.

\end{acknowledgments}

\appendix

\section{Dataset descriptions}
\label{sec:appendix-datasets}

Here we will give a more detailed descriptions of all datasets.  A
full description of each dataset can be found in the cited
references.  In some cases, the networks are created via complex
processes in special environments, so the true meaning of links and
groups may not have a simple interpretation.  Nevertheless, the
breadth of our data gives us a wide perspective on real-world, as
opposed to artificial, networks.

\texttt{lfr} - Lancichinetti-Fortunato-Radicchi benchmark graph with
1000 vertices ($N=1000$) and ``small'' communities (min size=10, max
size=50), at mixing parameter $\mu=0.5$ \cite{lancichinetti08}.  The
other parameters (average degree 20, maximum degree 50, exponent of
degree distribution -2, exponent of community size distribution -1)
are standard.  This graph has a clear community structure that is a
standard used to optimize and test most current algorithms, and thus
serves as a baseline reference for a network with known and detectable
structure.  The network was created with standard LFR code available
at \url{https://sites.google.com/site/santofortunato/inthepress2}.

\texttt{karate} - Karate club network.  A well known network of
friendships in a karate club in an American
University~\cite{zachary77}.  After a dispute between the coach and the
treasurer, the club split in two clubs. We use the standard unweighted
version, with two metadata groups defined by the membership after
the split.

\texttt{football} - American college football. Network of American football games between
Division IA colleges during the regular season Fall 2000~\cite{girvan02,Evans}.
Edges exist if two teams played any game, and groups are conferences,
scheduling groups joined by the schools for the purpose of regular season
scheduling.  Each season, conferences mandate and schedule a certain
number of intra-conference games played, and other matches are decided by
negotiation between schools. The data is available at
\url{http://www-personal.umich.edu/~mejn/netdata/}
contains conference assignments for year 2001, and an updated version (used in
this paper) contains correct conference assignments from Fall 2000, courtesy of
T.S. Evans\ \cite{Evans}.

\texttt{polblogs} - Political blogs.  A directed network of hyperlinks
between weblogs on US politics, recorded in 2005 by Adamic and
Glance~\cite{adamic05}.  Links are all front-page hyperlinks at the
time of the crawl. Groups are ``liberal'' or
``conservative'' as assigned by either blog directories or occasional
self-evaluation.  The data are available
at \url{http://www-personal.umich.edu/~mejn/netdata/}.

\texttt{polbooks} - Network of books about US politics from 2004 US
presidential election~\cite{krebs06} taken from the online bookseller
Amazon.com.  Edges are Amazon recommendations on each book, indicating
copurchasing by others on the site. Groups are based on
political alignment of ``liberal'', ``neutral'', or ``conservative''
through human evaluation.  Data can be found at
\url{http://www-personal.umich.edu/~mejn/netdata/}.  

\texttt{dpd} - Software dependencies within the Debian GNU/Linux
operating system~\cite{debian7.1, debianDevRef}.  Nodes are unique
software packages, such as \texttt{linux-image-2.6-amd64},
\texttt{libreoffice-gtk}, or \texttt{python-scipy}.  Links are the
``depends'', ``recommends'', and ``suggests'' relationships, which are
a feature of Debian's APT package management system designed for
tracking dependencies.  Groups are tag memberships from the
DebTags project, \url{https://wiki.debian.org/Debtags}, such as
\texttt{devel::lang:python} or \texttt{web::browser}~\cite{zini05}.
The network was generated from package files in Debian 7.1 Wheezy as of
2013-07-15, ``main'' area only.  Similar files are freely available in
every Debian-based OS.  Tags can be found in the \texttt{*\_Packages}
files in the \texttt{/var/lib/apt/} directory in an installed system
or on mirrors, for example
\url{ftp://ftp.debian.org/debian/dists/wheezy/main/binary-amd64/}.

\texttt{pgp} - The ``Web of trust'' of PGP (Pretty Good Privacy) key
signings, representing an indication of trust of the identity of one
person (signee) by another (signer)~\cite{garfinkel95}.  A node
represents one key, usually but not always corresponding to a real
person or organization.  Links are signatures, which by convention are
intended to only be made if the two parties are physically present,
have verified each others' identities, and have verified the key
fingerprints.  Groups are email
domain or subdomain names.  The network was generated using full data
downloaded from the \url{http://sks-keyservers.net} keyserver network.
Signatures were not checked for cryptographic validity.  Domains were
broken into all subdomains, for example the address
\texttt{example@becs.aalto.fi} would be added to the three groups
\texttt{becs.aalto.fi}, \texttt{aalto.fi}, and \texttt{fi}.  Large
webmail providers and top level domains were discarded by hand:
\texttt{com, info, net, org, biz, name, pro, edu, gov, int,
  gmail.com, yahoo.com, mail.com, excite.com, hotmail.com}

\texttt{as-caida} - Network of the Internet at the level of Autonomous Systems~\cite{rfc1930}.  Nodes represent
autonomous systems, i.e.\ systems of connected routers under the
control of one or more network operators with a common routing policy.
Links represent observed paths of Internet Protocol traffic
directly from one AS to another.  Groups are countries of
registration of each AS, which are by construction non-overlapping.
Data come from both the \textit{AS Relationships Dataset} from
2013-08-01~\cite{caida-as-rel} and \textit{The IPv4 Routed /24 AS Links
  Dataset} from 2013-01-01 to 2013-11-25~\cite{caida-ip-routed}.  This
means that our network contains every direct link observed by these
two subprojects on the Internet over a period of approximately one
year.  AS country assignments from all Regional Internet Registries
(AFRINIC, APNIC, ARIN, LACNIC, and RIPENCC) are taken from the mirror
\url{ftp://ftp.ripe.net/pub/stats/} on 2013-11-25.

%


\texttt{amazon} - Network of product copurchases on online retailer
Amazon.  Nodes represent products, and edges are said to represent copurchases by
other customers presented on the product page~\cite{amazon}.  The true meaning of links
is unknown and is some function of Amazon's recommendation algorithm.  Data was scraped in
mid-2006 and downloaded
from \url{http://snap.stanford.edu/data/amazon-meta.html}.  We used
copurchasing relationships as undirected edges. Product categories,
such as \texttt{Books/Fiction/Fantasy} or \texttt{Books/Nonfiction}
can be split into levels, which we used to make a fully hierarchical
network, for example \texttt{Books} in \texttt{layer00}, and
\texttt{Books/Fiction} and \texttt{Books/Nonfiction} in
\texttt{layer01}, down to \texttt{layer09}.  Finally, there is one
layer \texttt{categs} representing full categories, in this example
\texttt{Books/Fiction/Fantasy} and \texttt{Books/Nonfiction} even
though they contain a different number of ``\texttt{/}'' characters.

\texttt{anobii} - Social network of book recommendation, popular in
Italy. Two types of directed relationships were taken as undirected
links (“friends” and “neighbors”).  Users can form and join groups.
Data were provided by Luca Aiello~\cite{aiello12,aiello10}.

\texttt{dblp} - Network of collaboration of computer scientists. Two
scientists are connected if
they have coauthored at least one paper~\cite{backstrom06}. Groups are publication
venues (scientific conferences).  Data can be found at
\url{http://snap.stanford.edu/data/com-DBLP.html}~\cite{yang12}.




%
%

\texttt{fb100} - Facebook social networks.  100 complete (but
separate) Facebook networks at United States universities in 2005.
There are all friendships (undirected), as well as six pieces of node
metadata: dorm (residence hall), major, second major, graduation year,
former high school, and gender. These pieces of metadata were used to
form separate levels of groups. Networks were originally released
by M.\ A.\ Porter \cite{traud12} and are
available on several sites on the web.  The ``gender'' metadata were
discarded from the analysis as they form one giant network-spanning
group for male and female, with isolated fringes.

\texttt{flickr} - Picture sharing web site and social network, as
crawled by Alan Mislove~\cite{mislove07}.  Nodes are users and edges
exist if one user ``follows'' another.  Groups are Flickr user
groups centered around a certain type of content, such as
\texttt{Nature} or \texttt{Finland}.  The collectors estimate that
they have a vast majority of the LWCC by comparing to a random
sampling of users.  21\% of users are in groups.

\texttt{lj-backstrom} - LiveJournal social network, as crawled by Lars
Backstr\"om~\cite{backstrom06}.  The raw scrape from Livejournal, a
now-dormant blogging service. An edge was put between users if there
is any kind of relationship between them (friend or
follower). Groups are based on groups which users can join.

\texttt{lj-mislove} - LiveJournal social network, as crawled by Alan
Mislove~\cite{mislove07}.  The data source and node/edge/group
interpretation are the same as in \texttt{lj-backstrom}, but were
independently crawled.  61\% of users are in groups.

\texttt{orkut} - Orkut social network, as crawled by Alan
Mislove~\cite{mislove07}.  Nodes are users, edges are bidirectional
(undirected) friendships, and groups are user-created groups.
This crawl contains $10\%$ of Orkut's user population at the time of the
crawl (according to published figures).  Only $13\%$ of users are in groups.


\section{Community detection method descriptions}
\label{sec:appendix-methods}

This section contains a complete description of all community
detection methods and parameters used in this work.  Some methods do
not scale to the largest datasets, in which case 
results are not presented.  In analogy to the dataset preprocessing,
we also remove all detected communities of size less than 3.

\texttt{Infomap} (hierarchical mode) - Method
based on compression of the information associated to random walks on networks~\cite{rosvall11}.
Computed with code from \url{http://mapequation.org} with all default
settings.  

\texttt{InfomapSingle} (non-hierarchical mode) - Same as
\texttt{Infomap} but restricted to a non-hierarchical
partition~\cite{rosvall08}.  Computed with the same code as
\texttt{Infomap} but with the \texttt{--two-level}.

\texttt{Louvain} - Greedy, hierarchical modularity maximization
algorithm~\cite{blondel08}.  For each run, it is invoked 10 times and
the execution which has the maximal modularity (for each level) is taken.  The updated code from
\url{https://sites.google.com/site/findcommunities/} is used for
the calculations.

\texttt{Oslom} - Order Statistics Local Optimization Method, based on
community statistical significance~\cite{lancichinetti11}.  Code from
\url{http://oslom.org/} is used with all default settings, in
particular we run with 10 trials of the most granular level and 50
hierarchical trials of higher levels.  However, the \texttt{-singlet}
option is given, which causes all communities to be strictly
statistically significant.  Nodes not in any community are left as
singletons and then removed by our postprocessing, leaving community
assignments which do not cover the entire network.

\texttt{CliquePerc} - Clique percolation algorithm
from~\cite{reid12}.  Code from
\url{https://github.com/aaronmcdaid/MaximalCliques}.  We include one
layer for each clique size from $k=3$ to $k=k_\mathrm{max}$ for each
method.  By construction, each layer does not span the entire network.
Layers are numbered starting from layer $0$, which is percolation of cliques with
$k=3$ (triangles), up to layer $k_\mathrm{max}-3$, which is
percolation of cliques with $k=k_\mathrm{max}$.

\texttt{Copra} - A method based on label propagation~\cite{gregory10}.
Code from
\url{http://www.cs.bris.ac.uk/~steve/networks/software/copra.html} is
used with all default parameters.  In particular, this limits us to
non-overlapping communities with $v=1$, as there is no option for
automatically choosing the optimal parameter.

\texttt{Conclude} - A method using random walkers to re-weight
edges, then network distances are recalculated and used to
optimize weighted network modularity~\cite{demeo14}.  Code from
\url{http://www.emilio.ferrara.name/conclude/} is used with all
default options.

\texttt{Demon} - A method which combines node knowledge of local
neighborhoods into global communities~\cite{coscia12}.  Code from
\url{http://www.michelecoscia.com/?page_id=42} is used with all
default options.

\texttt{Ganxis} - Formerly the Speaker-listener Label Propagation
Algorithm (SLPA), a version of a label propagation
algorithm~\cite{xie12}.  Code version 3.0.2 from
\url{https://sites.google.com/site/communitydetectionslpa/} with
overlaps allowed, undirected mode, and one trial.  We chose all other default
parameters.  The code by default runs with eleven thresholds
$r \in \{0.01, 0.05, 0.1, 0.15, 0.2, 0.25, 0.3, 0.35, 0.4, 0.45,
0.5\}$, and all thresholds are kept (in order $0,\ldots,10$) in our
analysis.

\texttt{GreedyCliqueExp} - An algorithm, which finds
cliques as seeds and then optimizes a local fitness function around
those seeds~\cite{lee10}.  Code from
\url{https://sites.google.com/site/greedycliqueexpansion/} (version
r2011-11-06) is used with all default parameters.

\texttt{LinkCommunities} - Method partitioning
links, instead of nodes, into communities~\cite{ahn10}.  Code from
\url{http://barabasilab.neu.edu/projects/linkcommunities/} is used
with all default parameters.  Instead of scanning all thresholds, we
use three thresholds: $0.25$, $0.5$, and $0.75$ which are identified
as layer 0, layer 1, and layer 2 respectively.  All default parameters
are kept.  Links which are not part of any community at a given
threshold become \textit{singleton links}, which become communities of
size two.  These communities have no significance, and thus are
filtered out in our postprocessing.

\section{NMI grids}

\begin{figure*}
  \def\widthA{0.5\textwidth}
  \centerline{
  \begin{tabular}{cc}
    \includegraphics[width=\widthA]{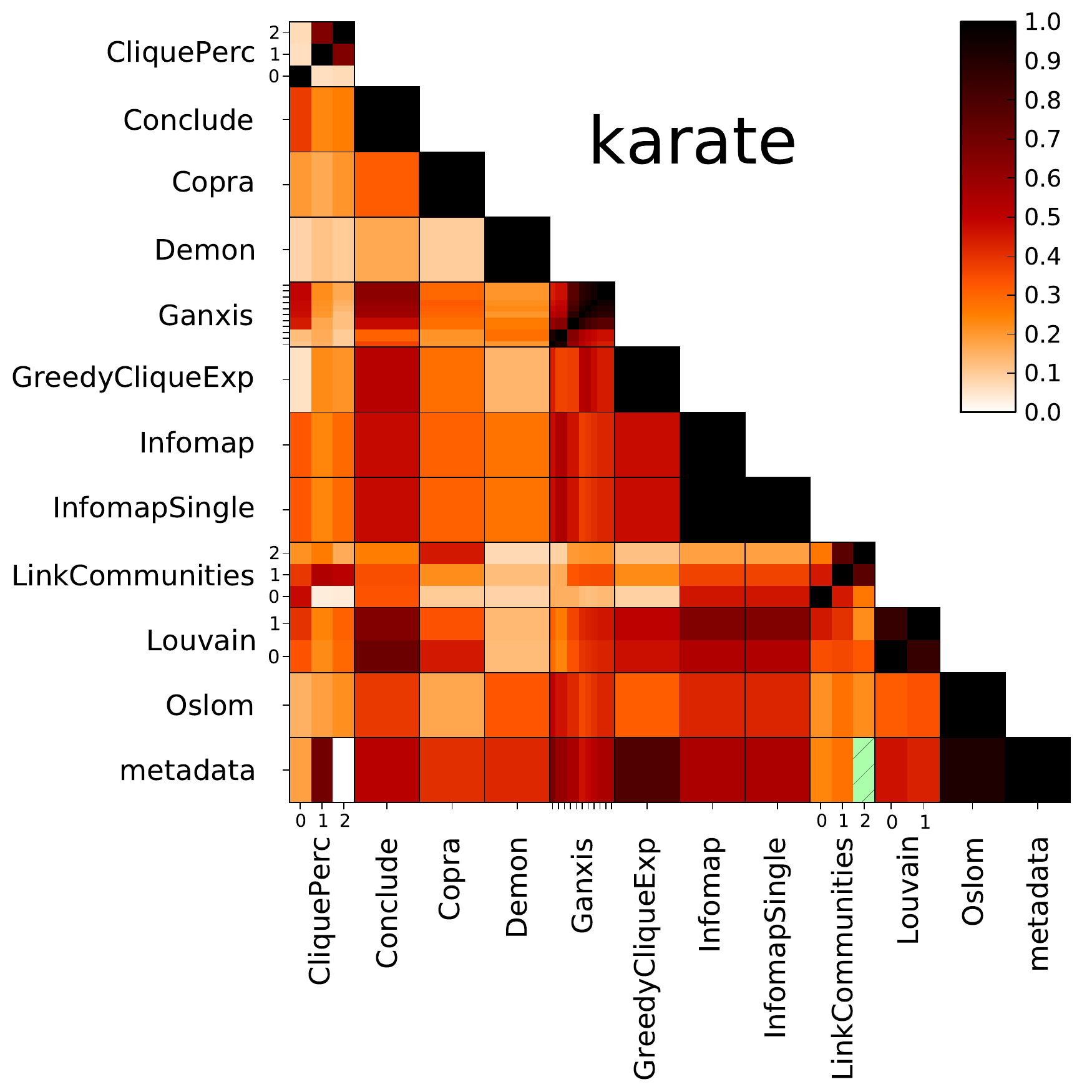} &
    \includegraphics[width=\widthA]{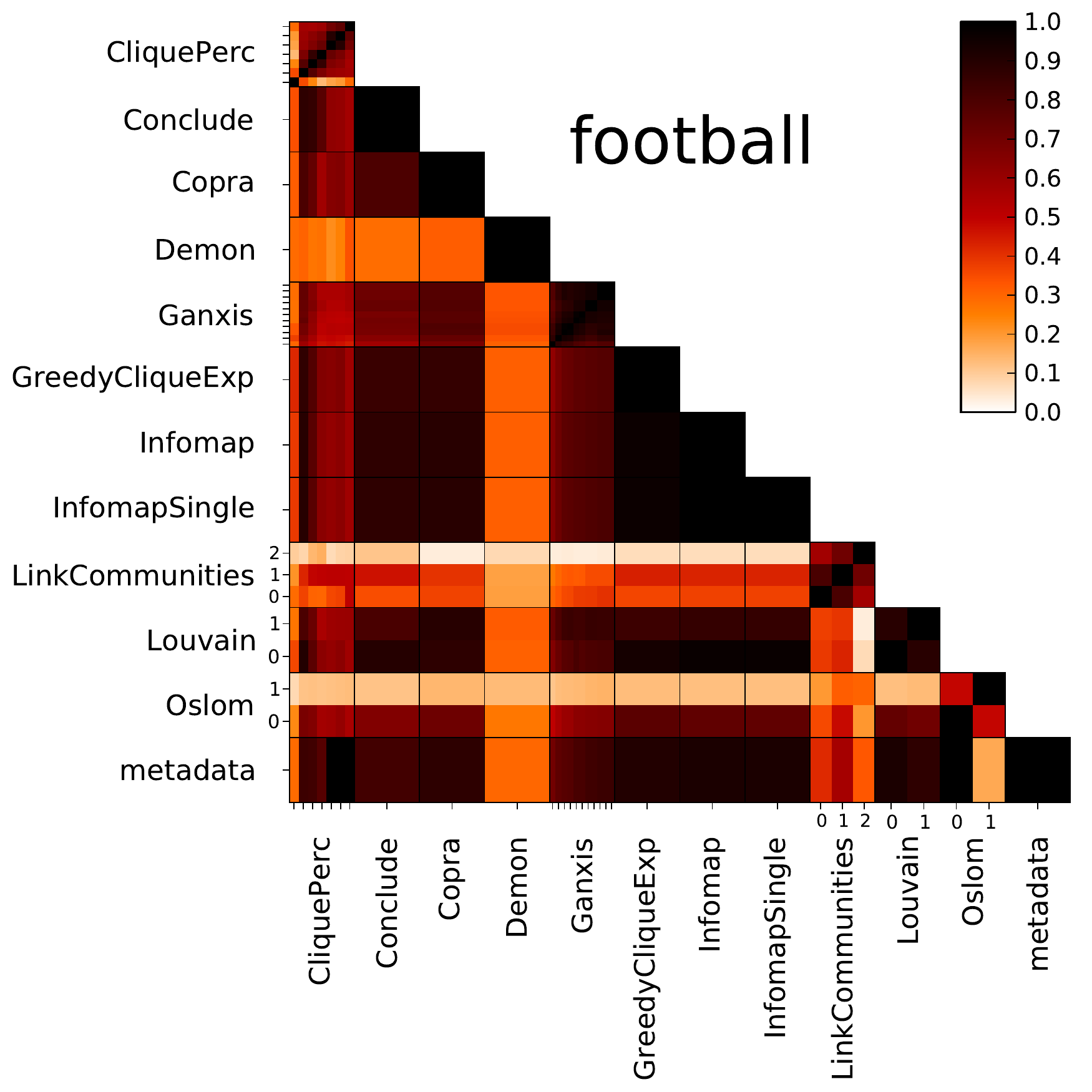} \\
    \includegraphics[width=\widthA]{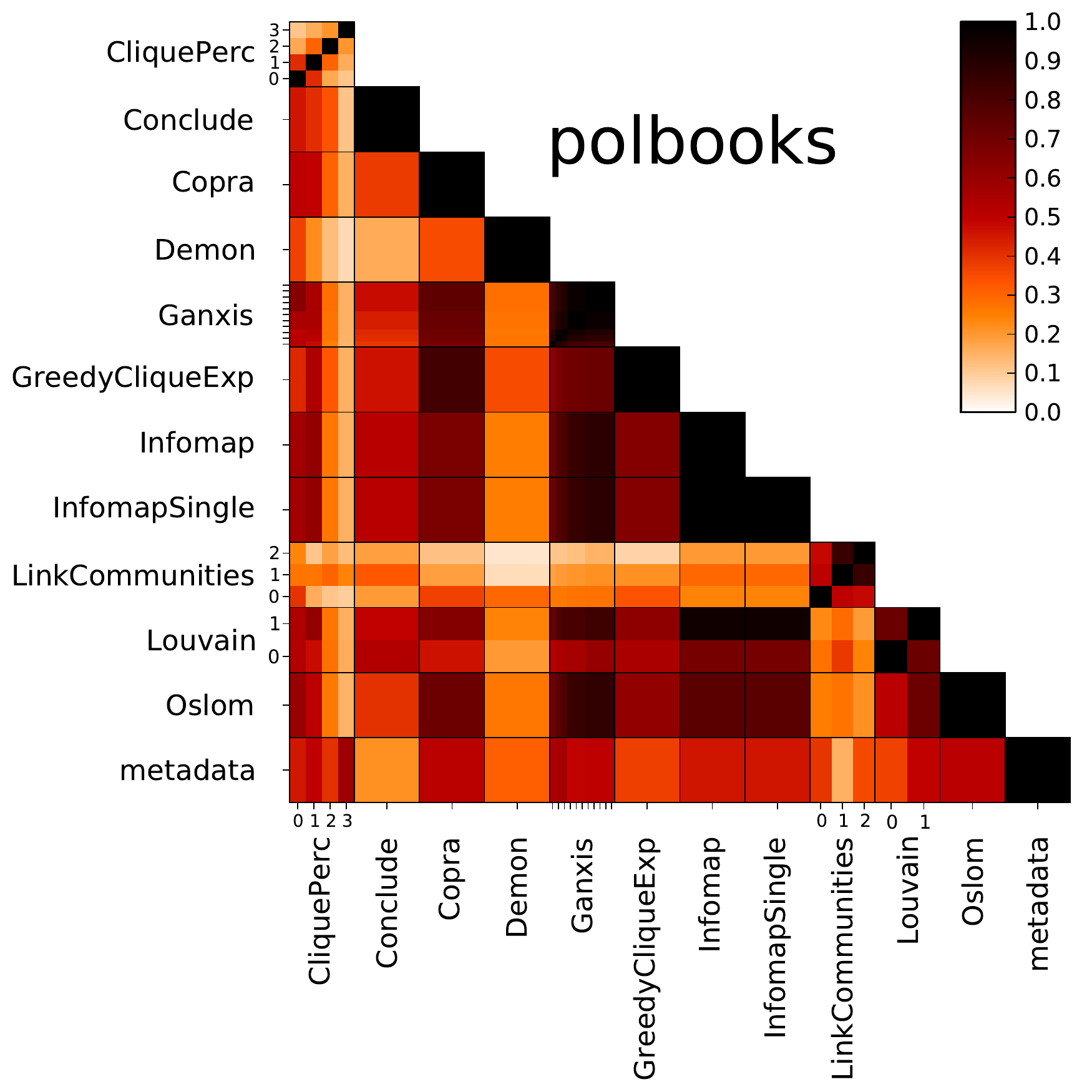} &
    \includegraphics[width=\widthA]{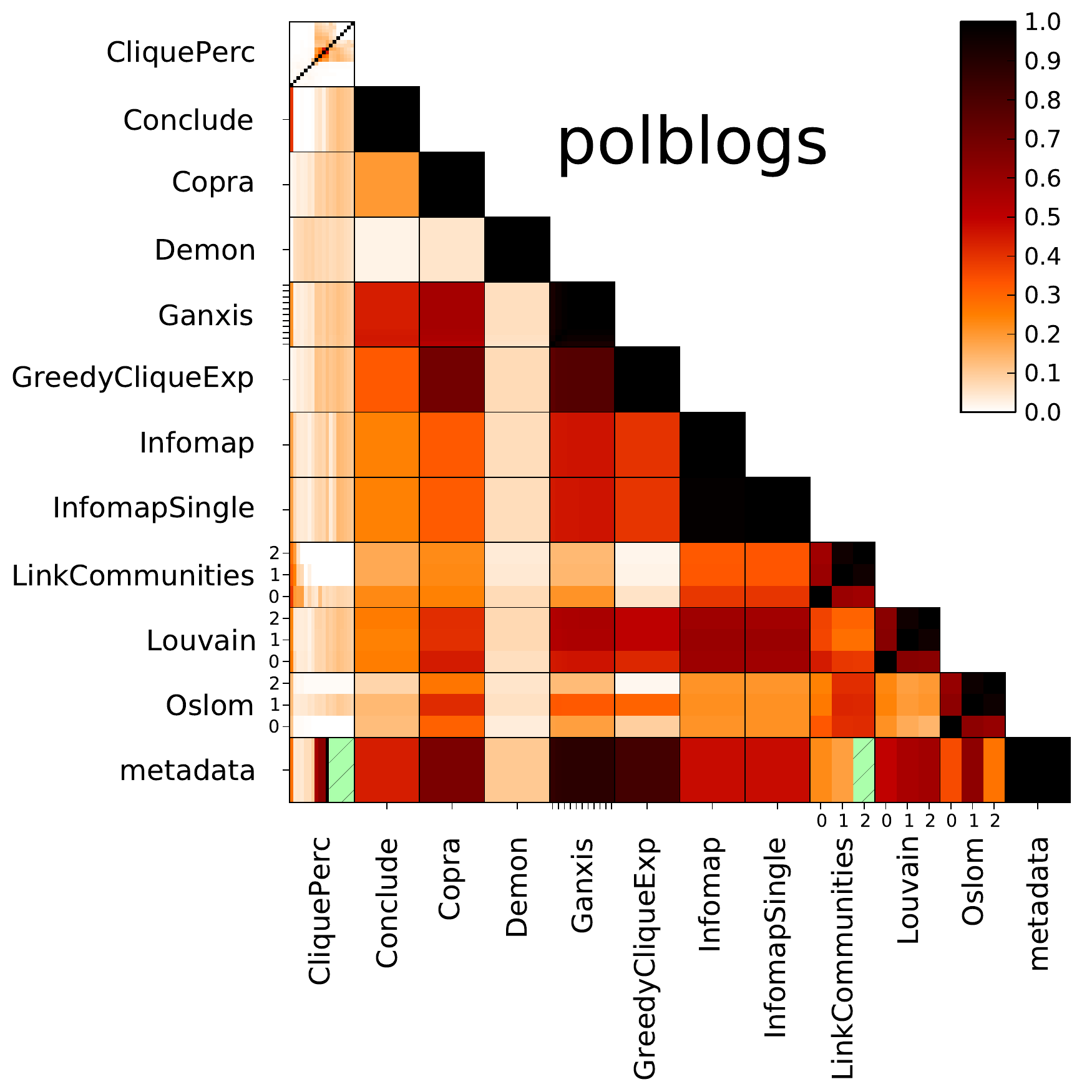}
  \end{tabular}}
  \caption{(Color online) NMI grids of \texttt{karate, football, polbooks} and
\texttt{polblogs}. These datasets have a pronounced community structure, which is the
reason why they are heavily used in the community detection literature. The
algorithms are quite successful in detecting the metadata groups (bottom row) and the
cross algorithmic stability is quite good, although still lower than expected.
  }
  \label{fig:nmi_grid_0}
\end{figure*}

\begin{figure*}
  \def\widthA{0.5\textwidth}
  \centerline{
  \begin{tabular}{cc}
    \includegraphics[width=\widthA]{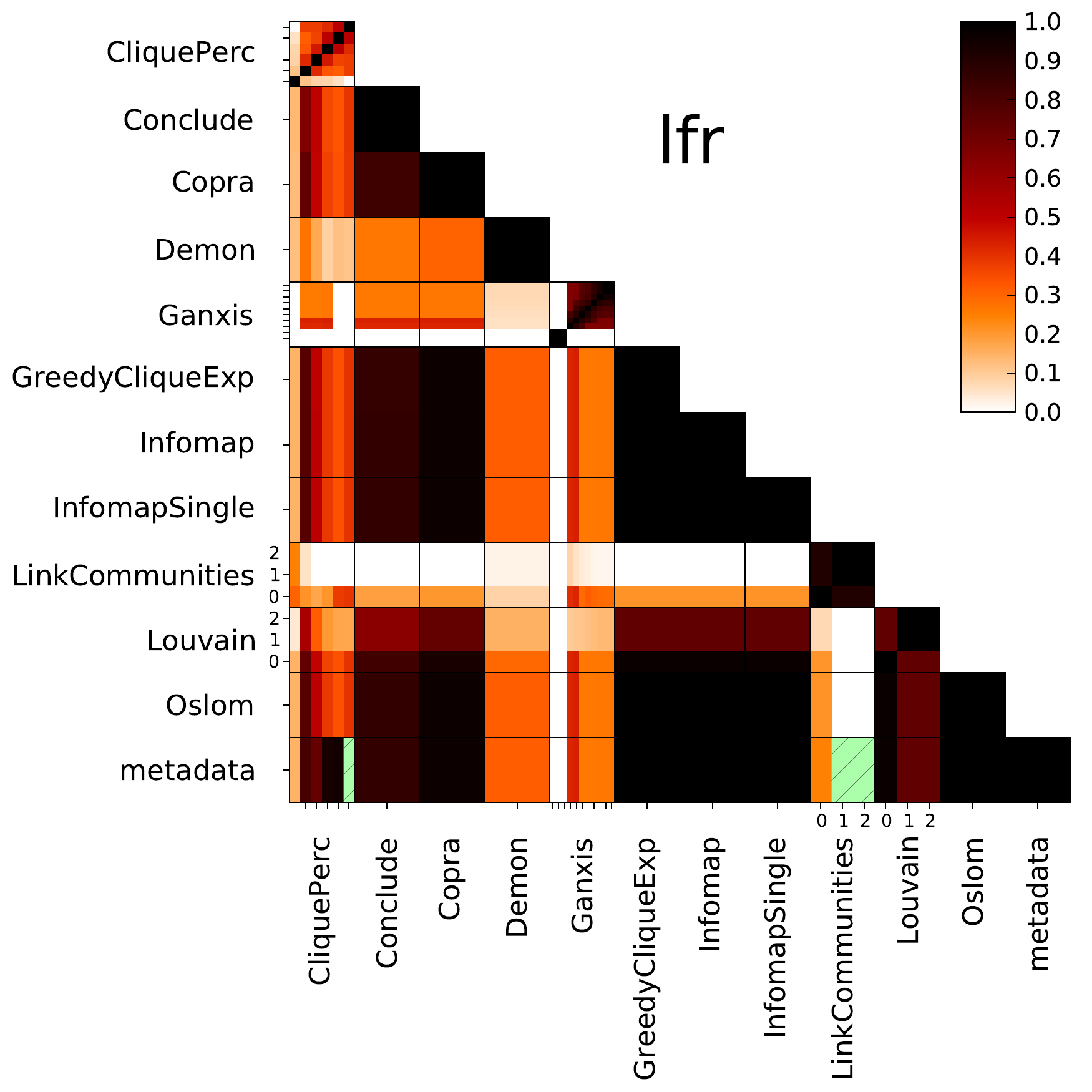} &
    \includegraphics[width=\widthA]{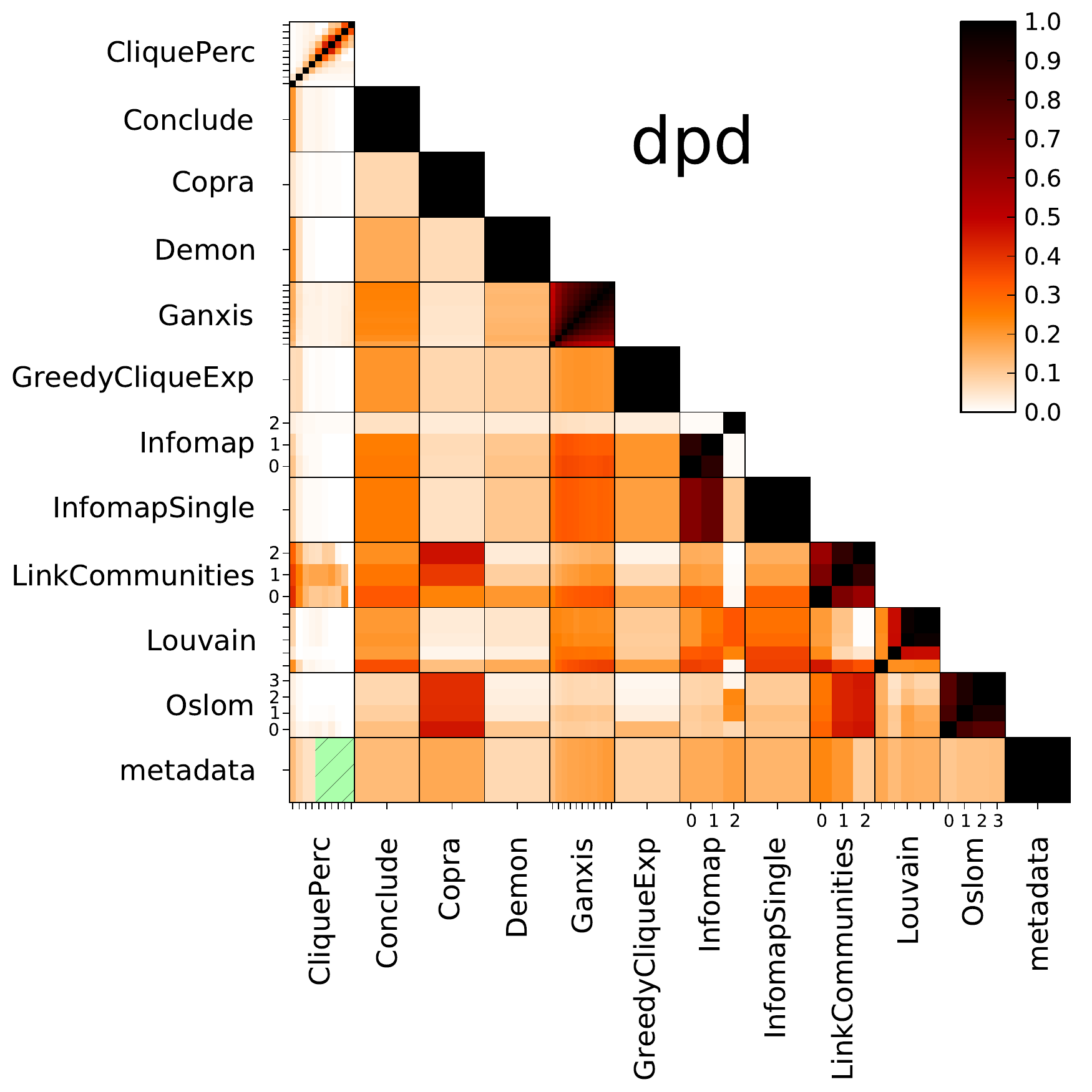} \\
    \includegraphics[width=\widthA]{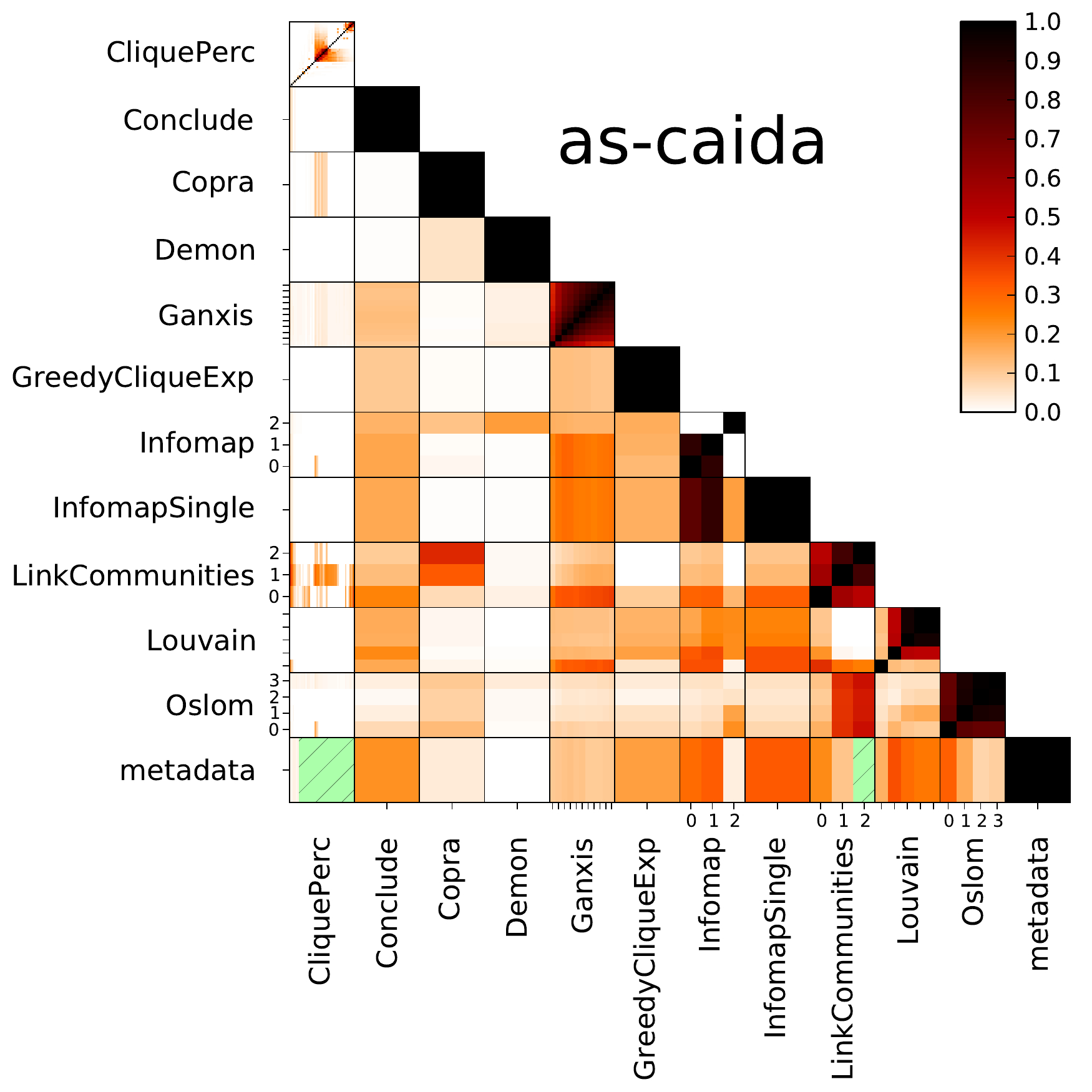} &
    \includegraphics[width=\widthA]{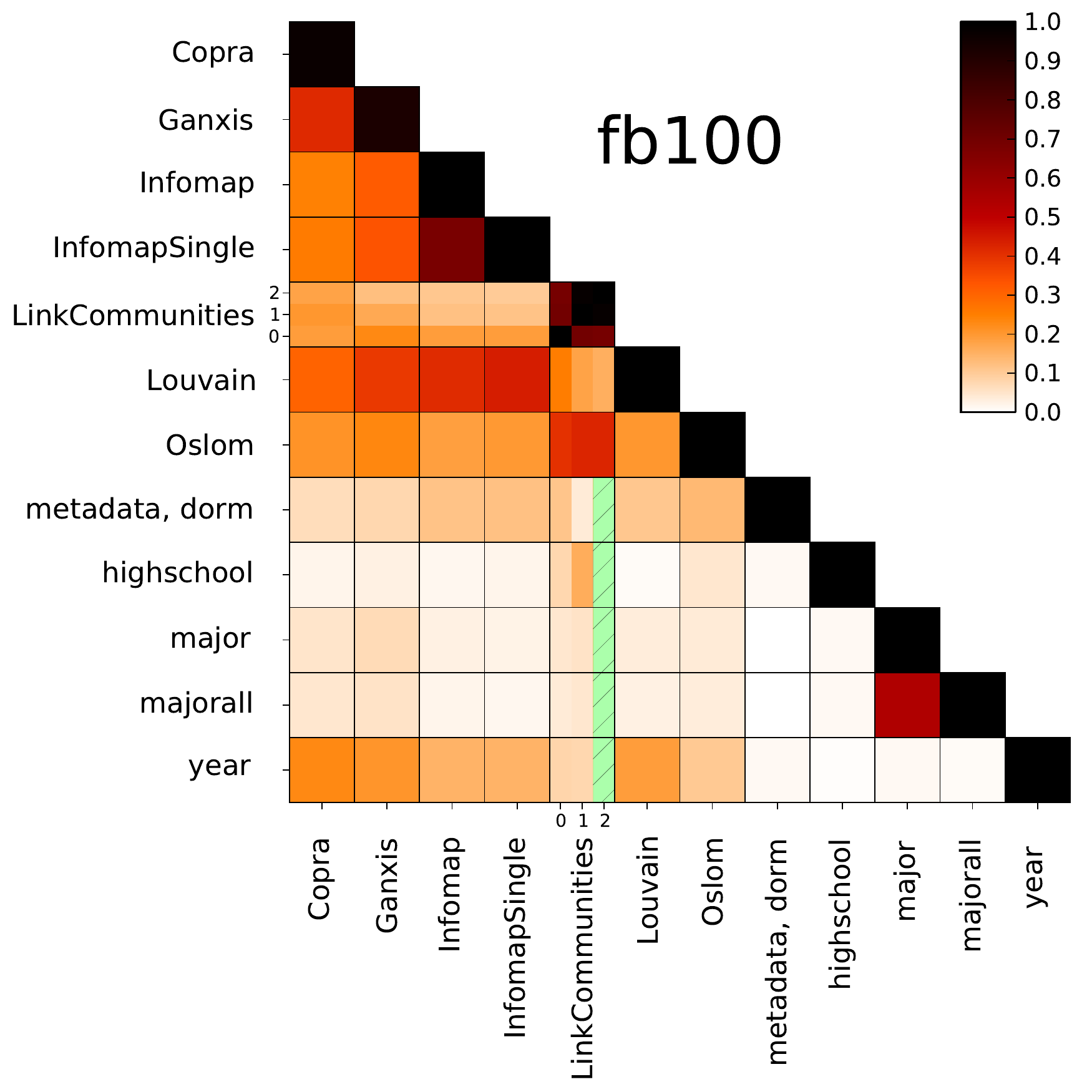}
  \end{tabular}}
  \caption{(Color online) NMI grids of \texttt{lfr, dpd, pgp} and \texttt{as-caida}. The first
dataset (\texttt{lfr}) is computer-generated.
Some algorithms performed poorly, whereas others scored very well. 
Other datasets are taken from the real world, and were a bigger
challenge. \texttt{fb100} is a collection of 100 datasets, so we report
the averaged maximum values for each tile, except for \texttt{LinkCommunities}, where the
number of layers is fixed at 3. Groups built using graduation year were
detected the best.
  }
  \label{fig:nmi_grid_1}
\end{figure*}

\begin{figure*}
  \def\widthA{0.5\textwidth}
  \centerline{
  \begin{tabular}{cc}
    \includegraphics[width=\widthA]{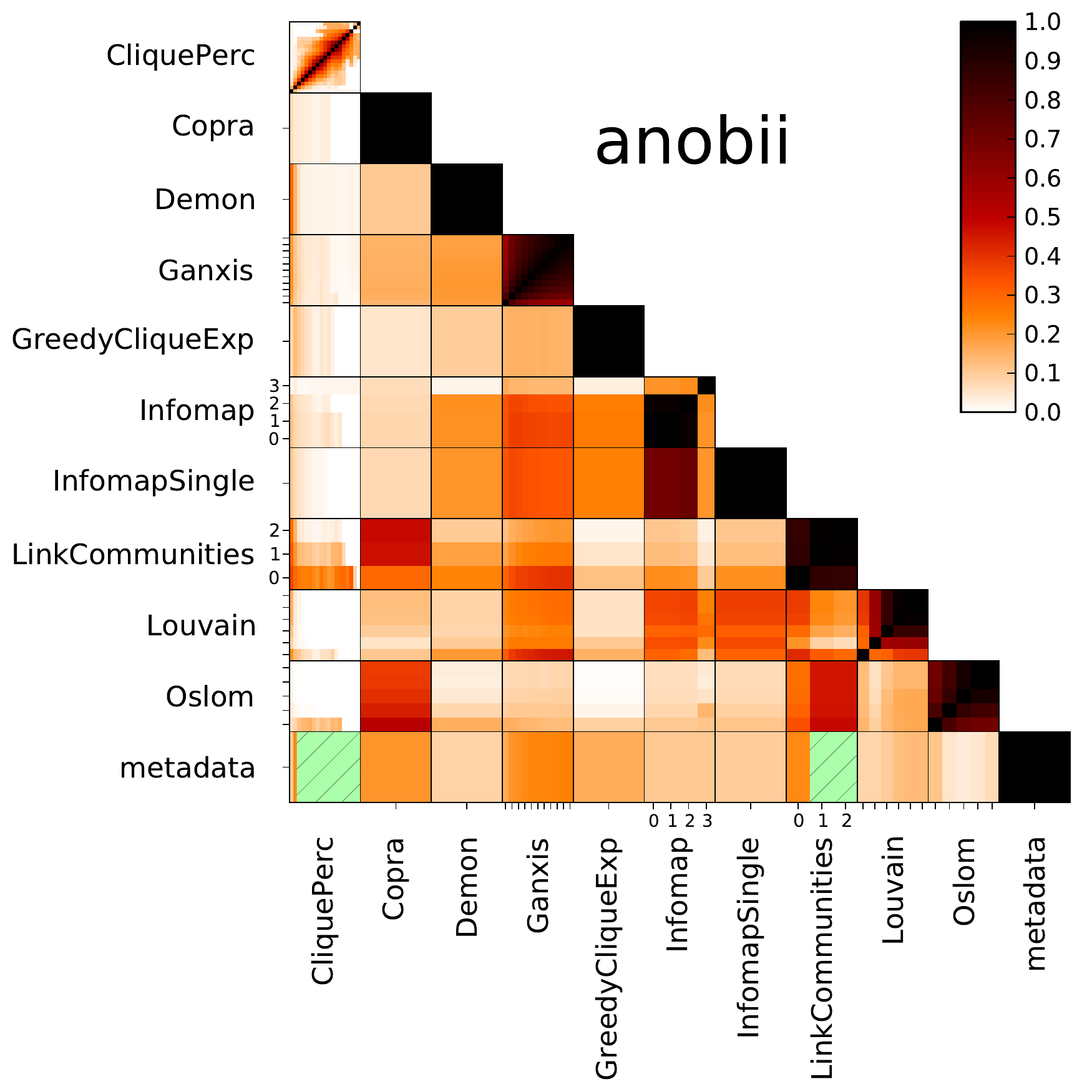} &
    \includegraphics[width=\widthA]{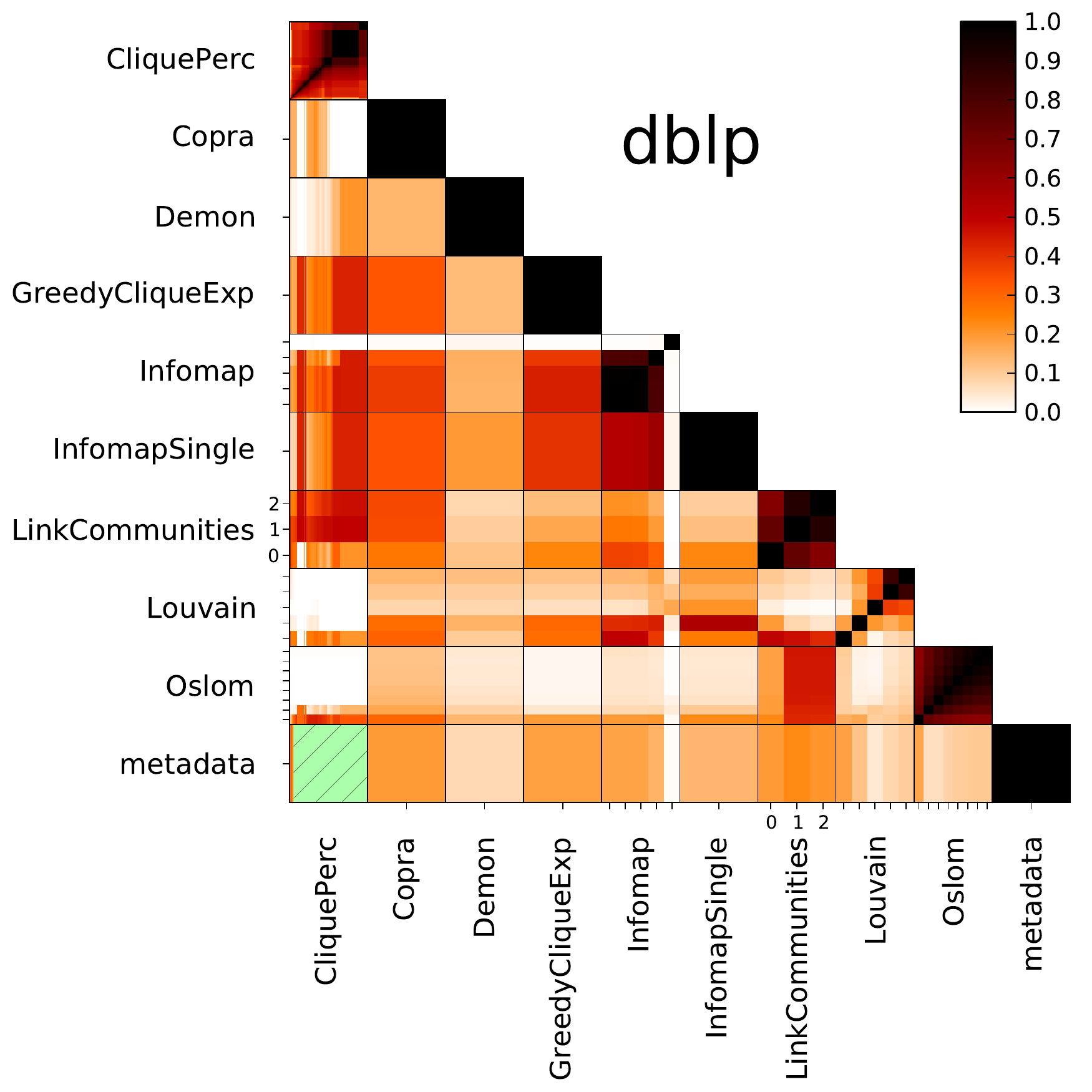} \\
    \includegraphics[width=\widthA]{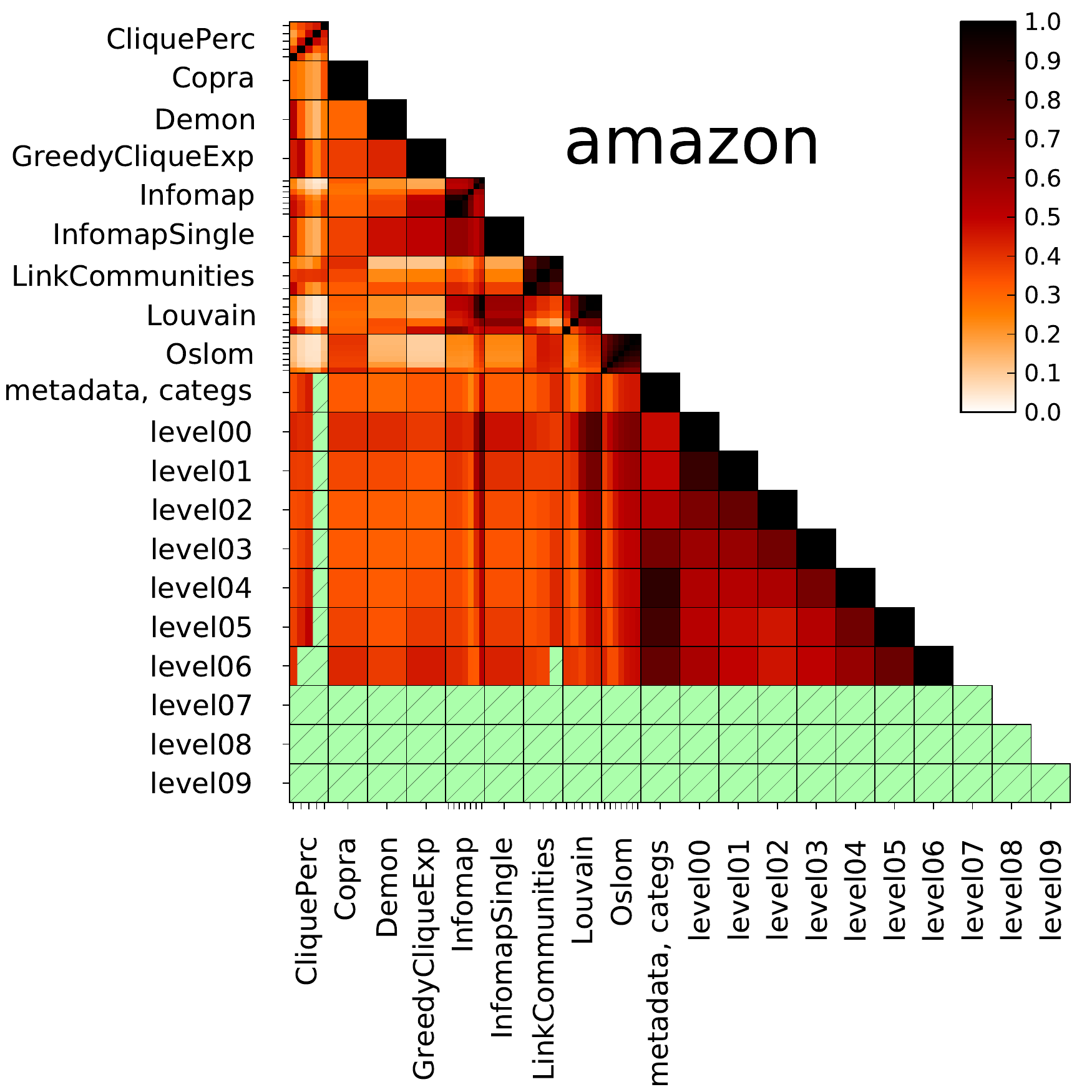}
  \end{tabular}}
  \caption{(Color online) NMI grids of \texttt{anobii, dblp} and \texttt{amazon}. \texttt{CliquePerc}
returned many spurious layers for \texttt{anobii} and \texttt{dblp}, that were
discarded due to poor coverage. More can be told for\texttt{amazon},
which contains hierarchical levels of product categories as different levels.
Deeper levels were discarded, but higher ones are detected, to some degree.
  }
  \label{fig:nmi_grid_2}
\end{figure*}

\begin{figure*}
  \def\widthA{0.5\textwidth}
  \centerline{
  \begin{tabular}{cc}
    \includegraphics[width=\widthA]{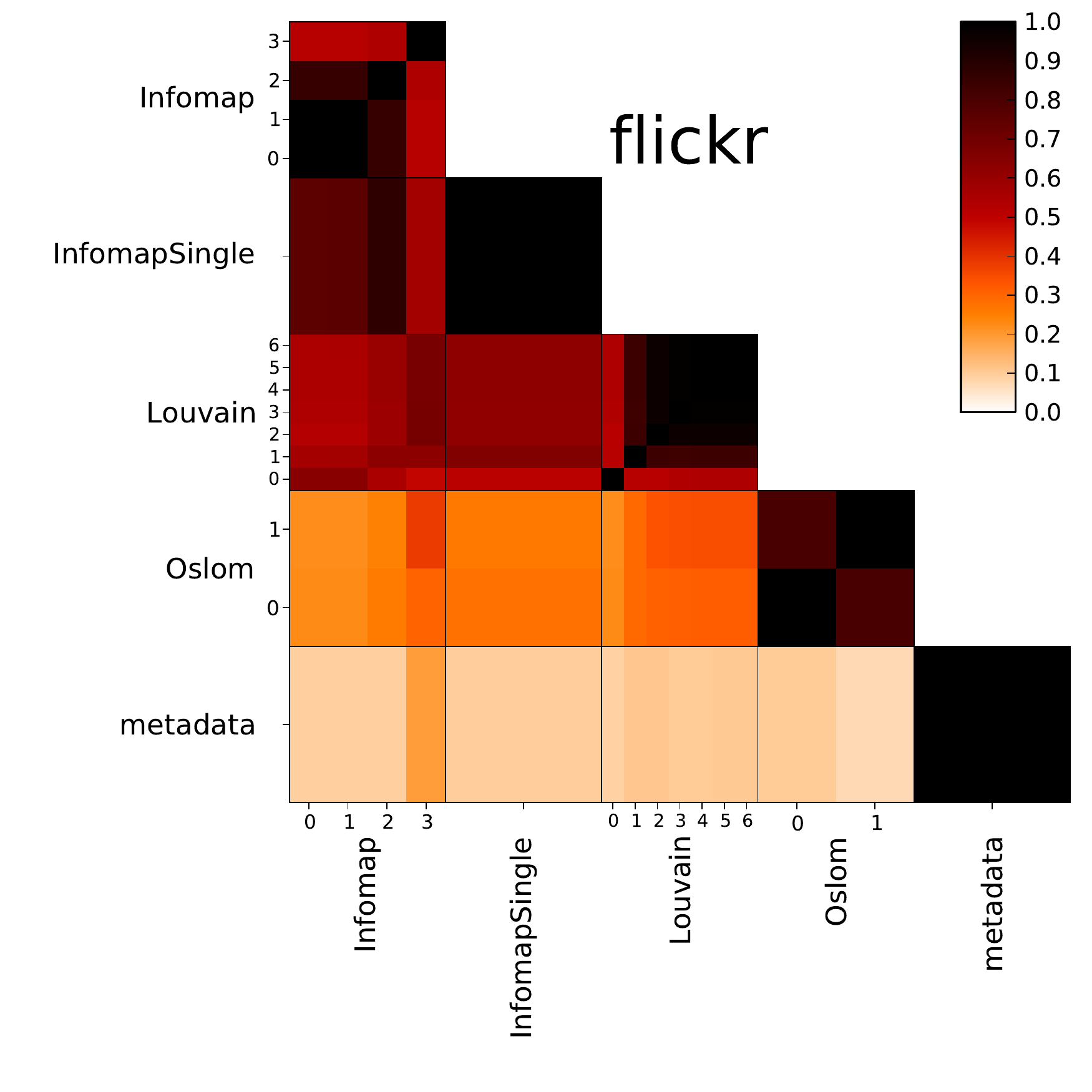} &
    \includegraphics[width=\widthA]{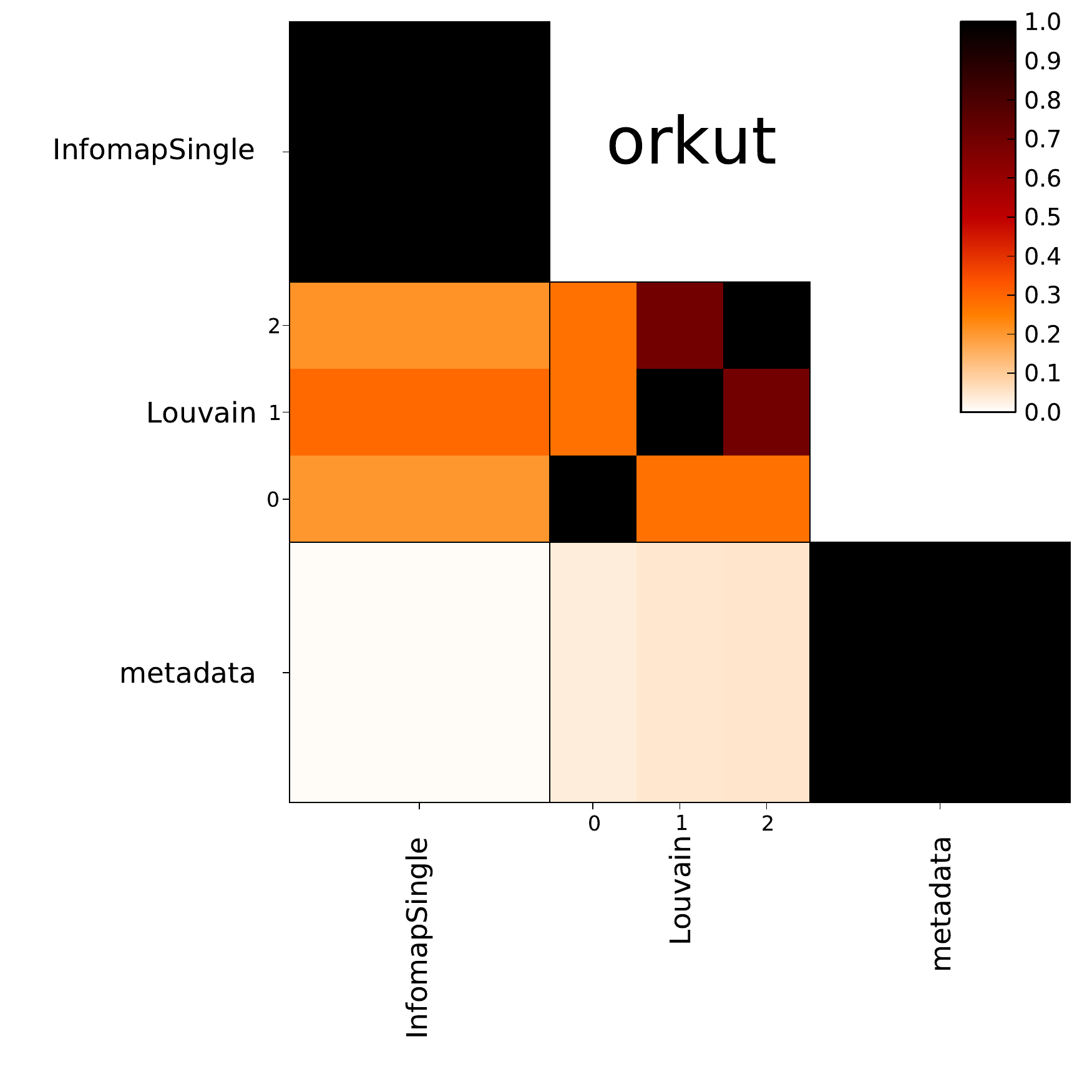} \\
    \includegraphics[width=\widthA]{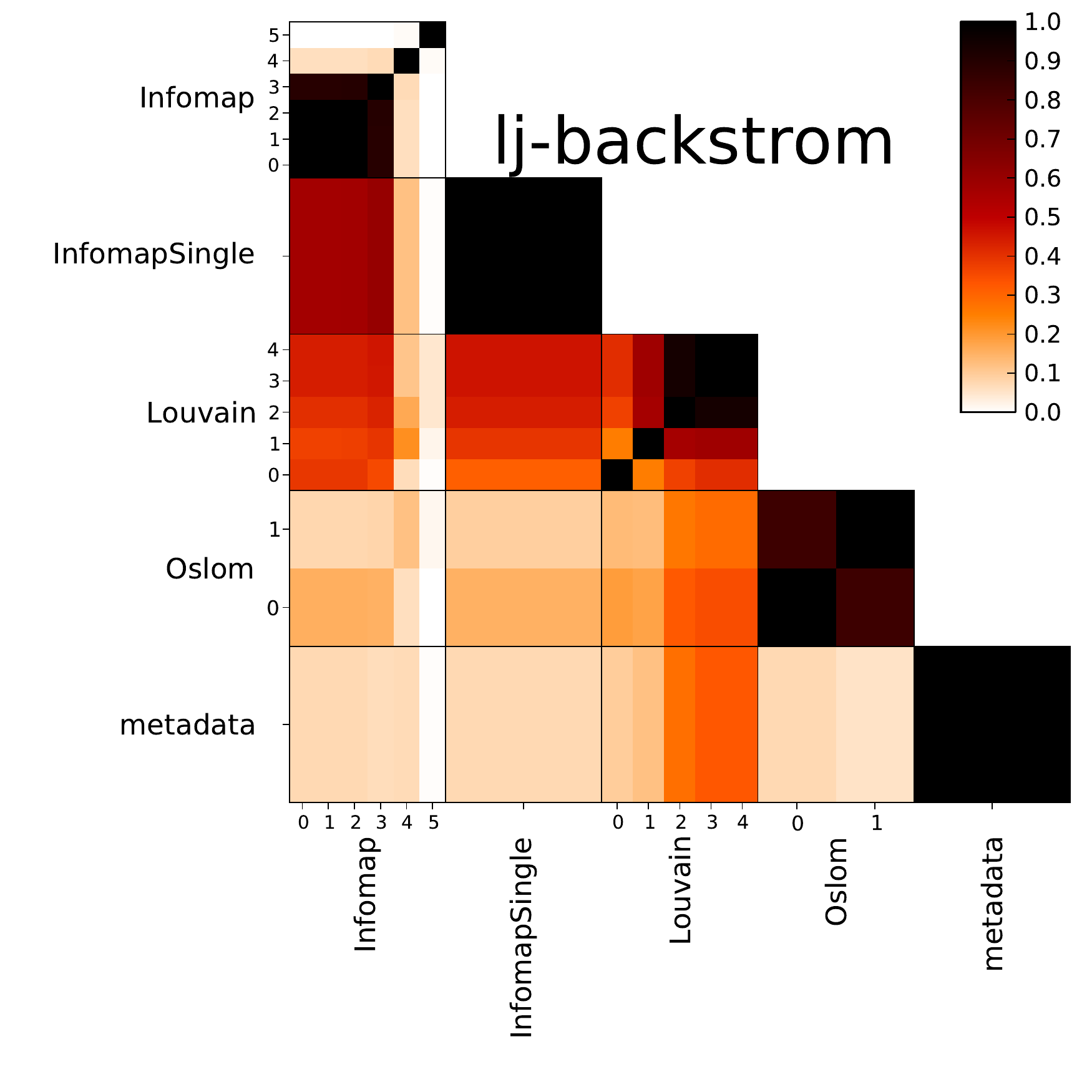} &
    \includegraphics[width=\widthA]{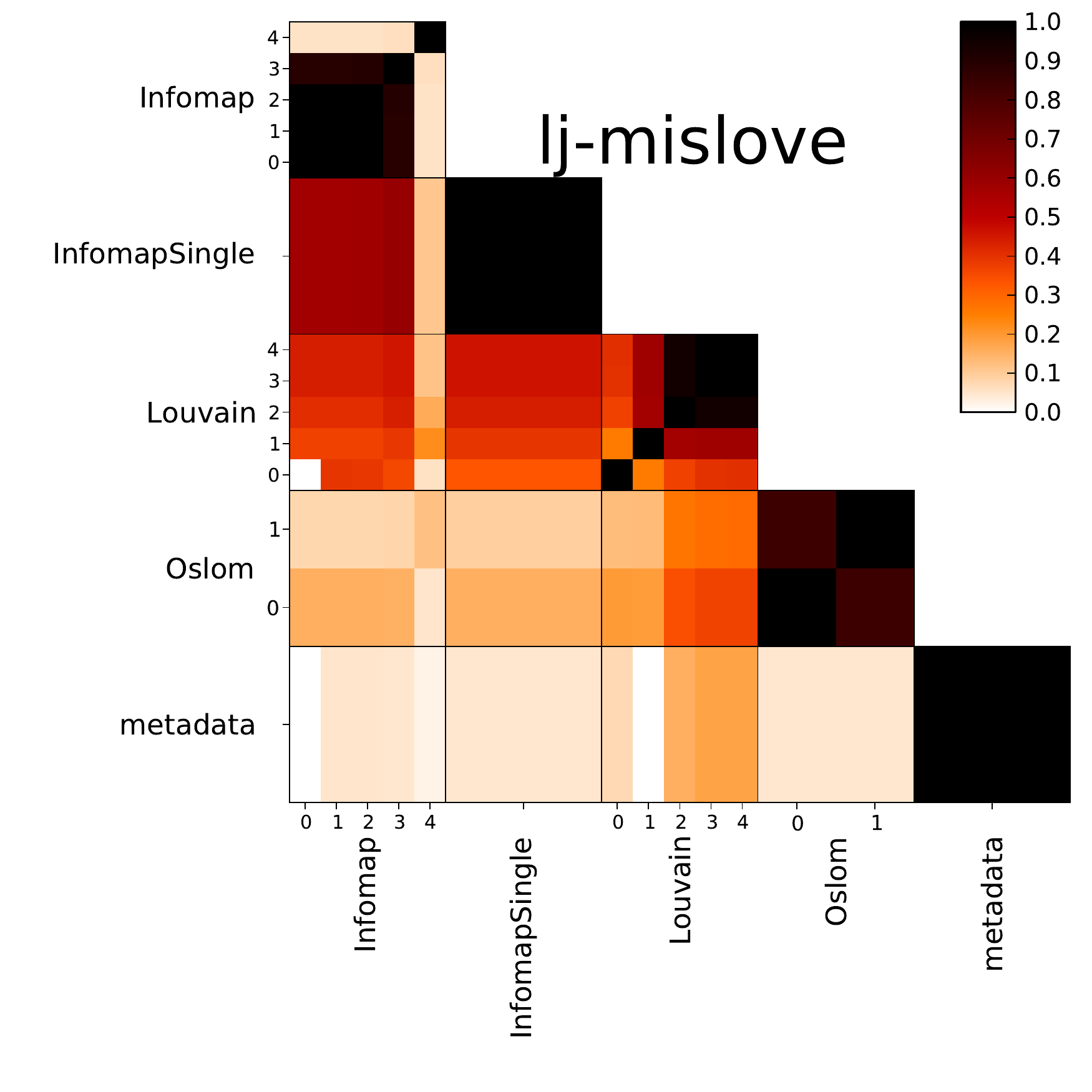}
  \end{tabular}}
  \caption{(Color online) NMI grids of \texttt{flickr, orkut, lj-backstrom} and
\texttt{lj-mislove}. Due to their size, many algorithms could not be run on
these datasets. The detection of the metadata partitions was poor, while the
similarity of detected partitions is noticeably higher. This suggests that
there is a large disparity between metadata and topological groups.
  }
  \label{fig:nmi_grid_3}
\end{figure*}

Here we present NMI grids for all datasets,
Figs.~\ref{fig:nmi_grid_0},~\ref{fig:nmi_grid_1},~\ref{fig:nmi_grid_2},
~\ref{fig:nmi_grid_3}.
The description is the same as for
\texttt{pgp} in Section~\ref{section:pgp}.

Most of the higher order layers of \texttt{CliquePerc} (and
suboptimal threshold parameter values in \texttt{LinkCommunities}),
after removing singletons and doubletons, cover a very small portion
of each dataset (less than $10\%$ of the nodes), and are marked with hatched
green. Larger datasets lack the results of the slowest algorithms due to computational restrictions.

\section{Additional community-level analysis}
\label{sec:appendix-jaccard}

In Section~\ref{sec:jaccard}, we showed that when matching
community-to-community, communities detected by various algorithms
often do not correspond to ``true'' groups, or vice versa.  In
this section, we will further this analysis to show that there is
little opportunity for narrowing our scope to increase the predictive
power of community detection methods.  



\def\kin{\ensuremath{k_\mathrm{in}}}
\def\kout{\ensuremath{k_\mathrm{out}}}
\def\ktot{\ensuremath{k_\mathrm{tot}}}

\begin{figure*}
\centering
\begin{minipage}[t]{\columnwidth}
  \centering
  \includegraphics[width=\columnwidth]{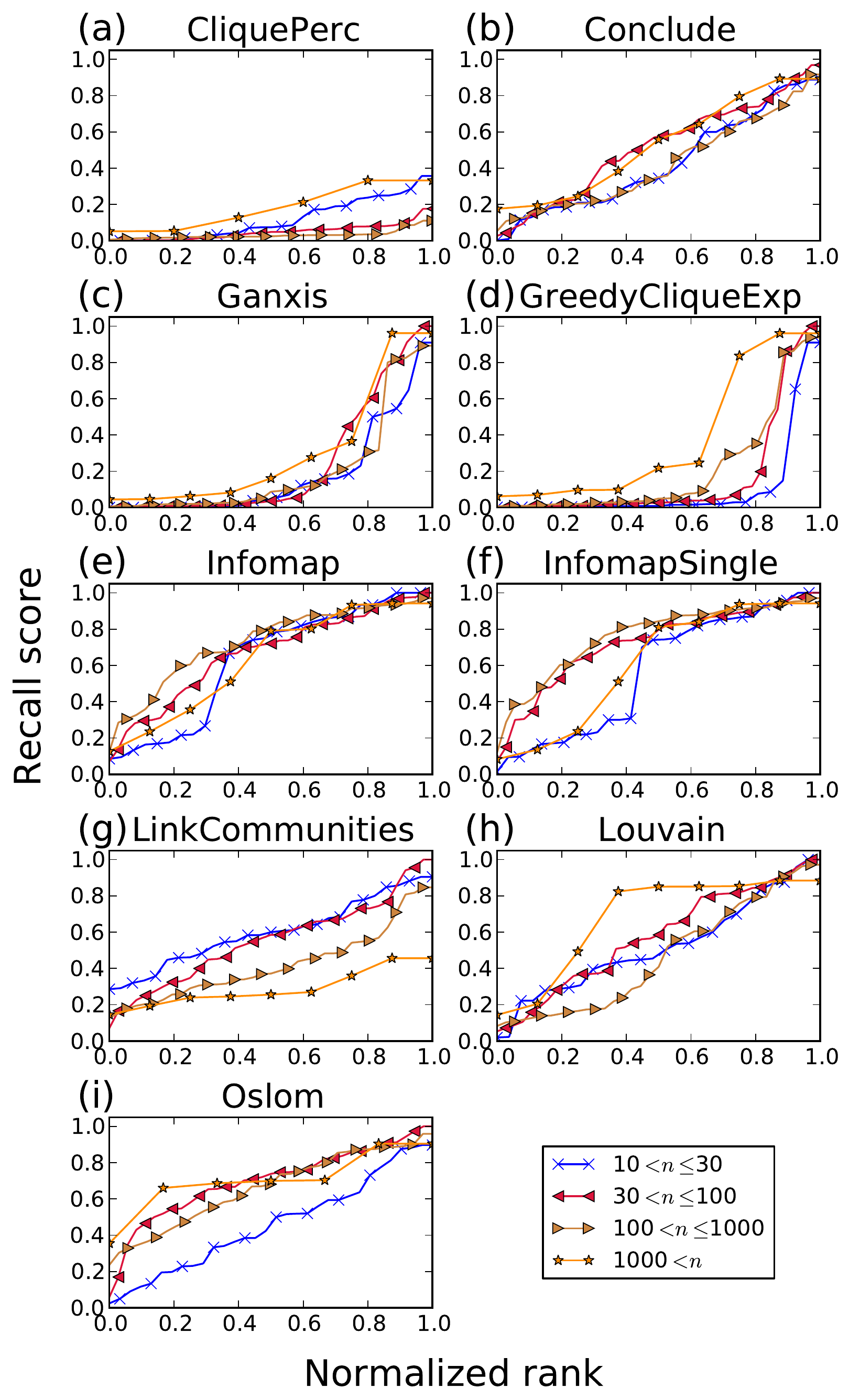}
  \caption{(Color online).  Recall of known \DS{as-caida} groups,
    broken down by size $n$ (of known groups), matched to all
    detected communities.  We see that most methods do not have a good
    performance for most group sizes.  }
  \label{fig:as-caida-recl-size}
\end{minipage}\hfill
\begin{minipage}[t]{\columnwidth}
  \centering
  \includegraphics[width=\columnwidth]{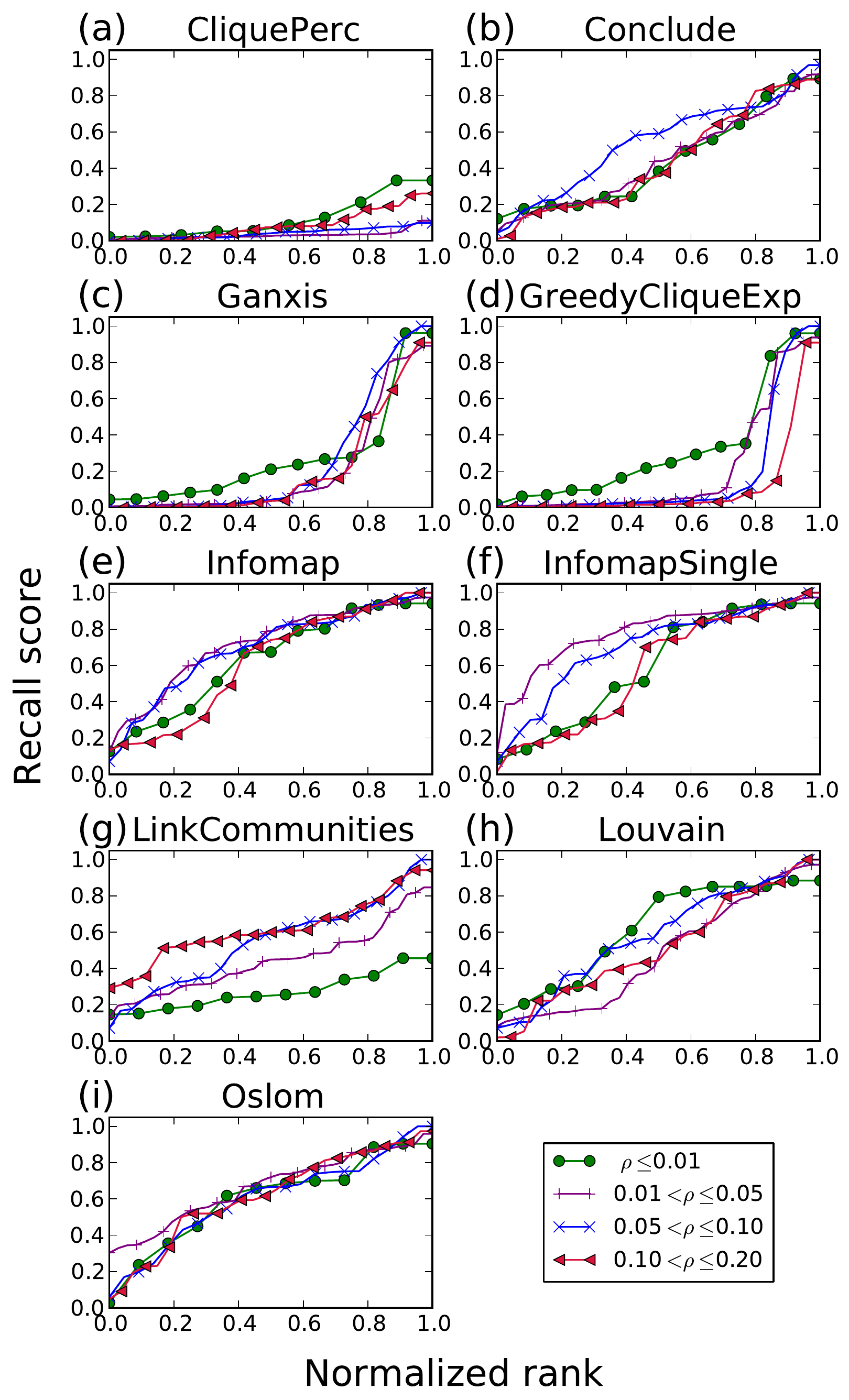}
  \caption{(Color online).  Recall of known \DS{as-caida}
    groups, broken down by density, matched to all detected
    communities.  Density is heavily correlated with inverse size,
    explaining the apparently higher performance on less dense
    groups.}
  \label{fig:as-caida-recl-den}
\end{minipage}
\end{figure*}

\begin{figure*}
\centering
\begin{minipage}[t]{\columnwidth}
  \centering
  \includegraphics[width=\columnwidth]{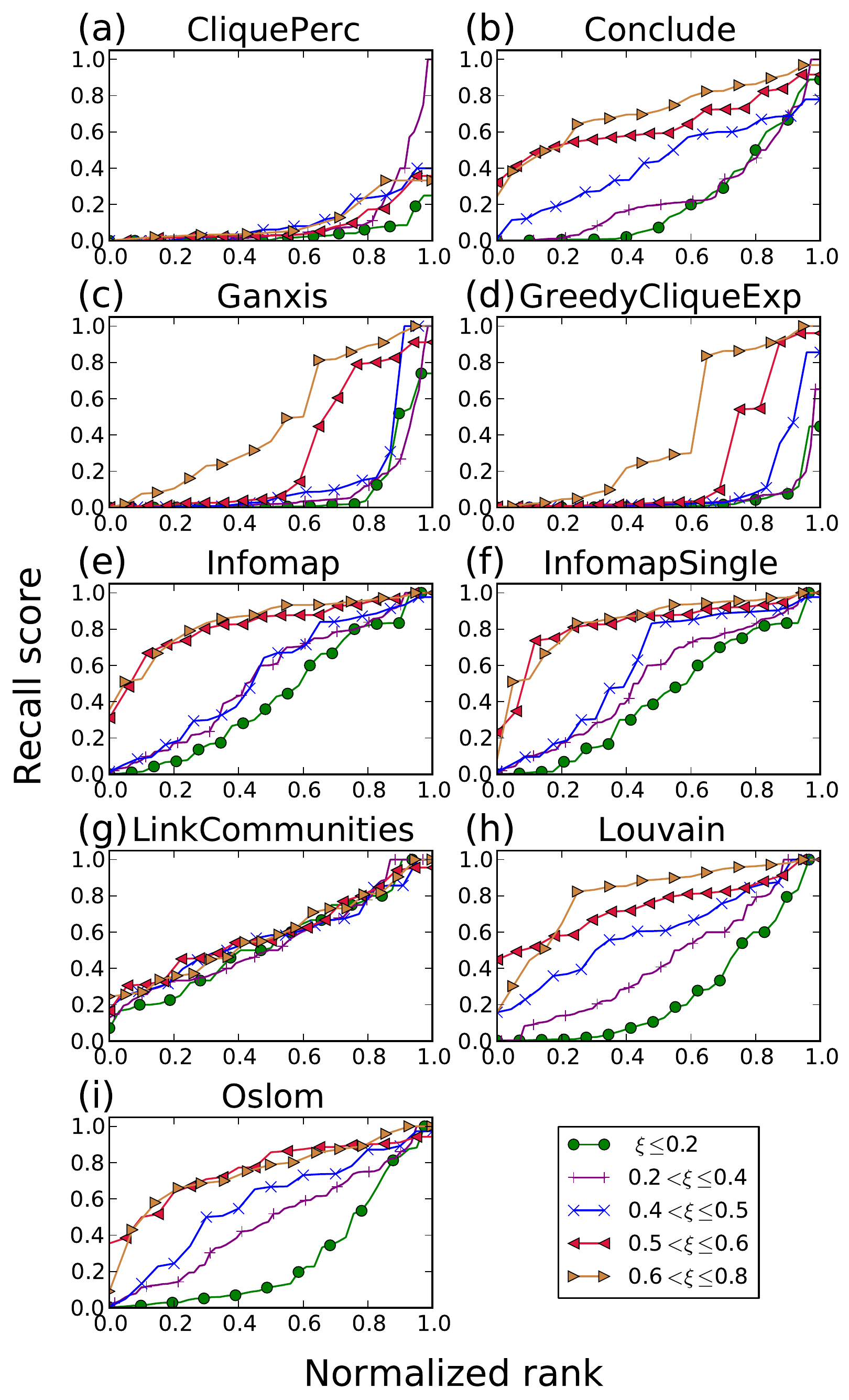}
  \caption{(Color online).  Recall of known \DS{as-caida} groups,
    broken down by group embeddedness $\xi=\kin/\ktot$,
    matched to all detected communities.  We see that, unlike size and
    density, embeddedness can very well predict the detectability of
    metadata groups.  Higher embeddedness directly corresponds
    to better detectability for almost all methods.}
  \label{fig:as-caida-recl-ce}
\end{minipage}\hfill
\begin{minipage}[t]{\columnwidth}
  \centering
  \includegraphics[width=\columnwidth]{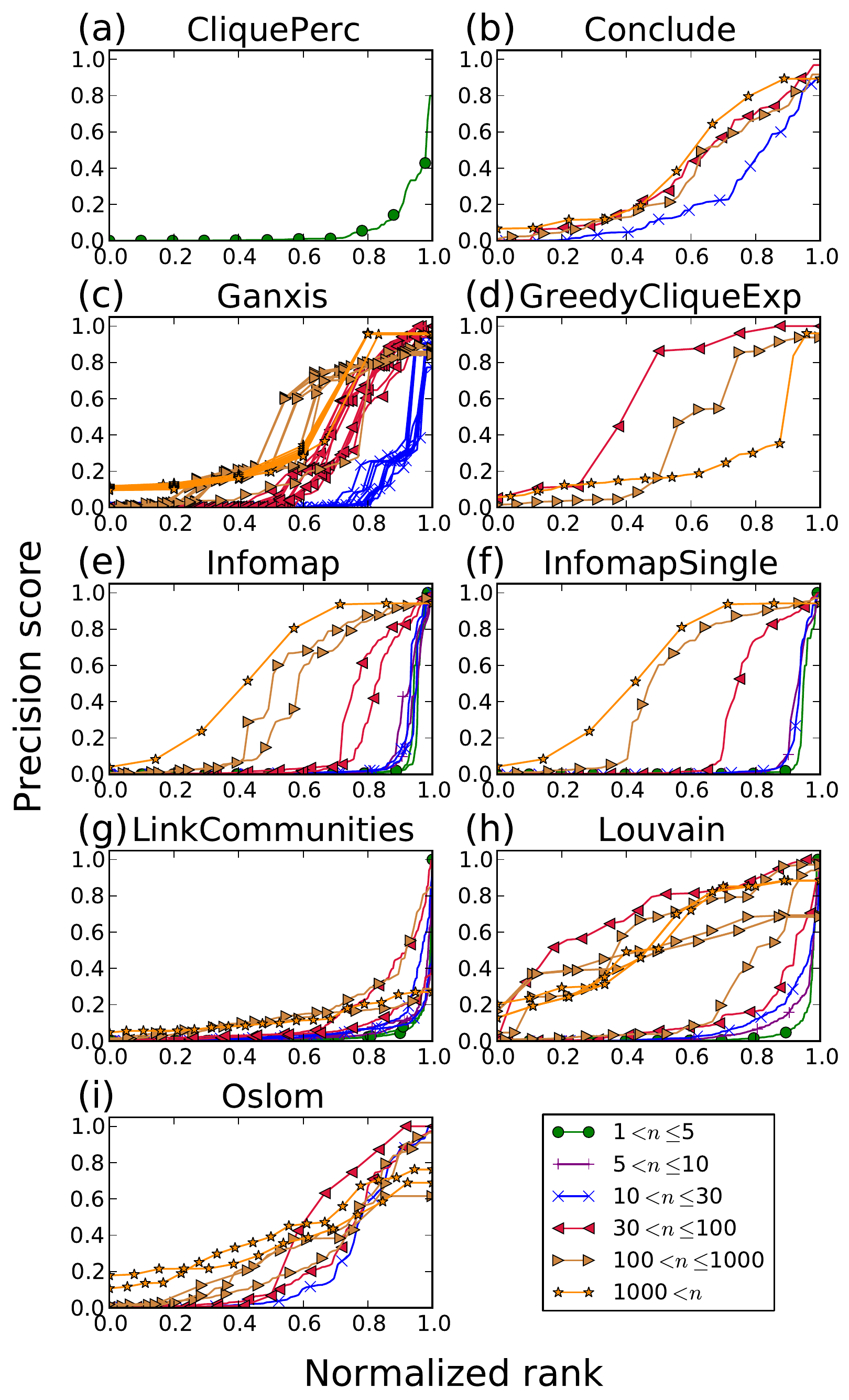}
  \caption{(Color online).  Precision of detected communities, broken
    down by size, compared to all known \DS{as-caida} groups.
    The recall of each layer of the algorithm is plotted separately to
    allow us to see if any individual layers have high performance.
    This produces a very messy field of lines, but it is sufficient to
    see that there are no outliers in performance.}
  \label{fig:as-caida-prec-size}
\end{minipage}
\end{figure*}

\begin{figure*}
\centering
\begin{minipage}[t]{\columnwidth}
  \centering
  \includegraphics[width=\columnwidth]{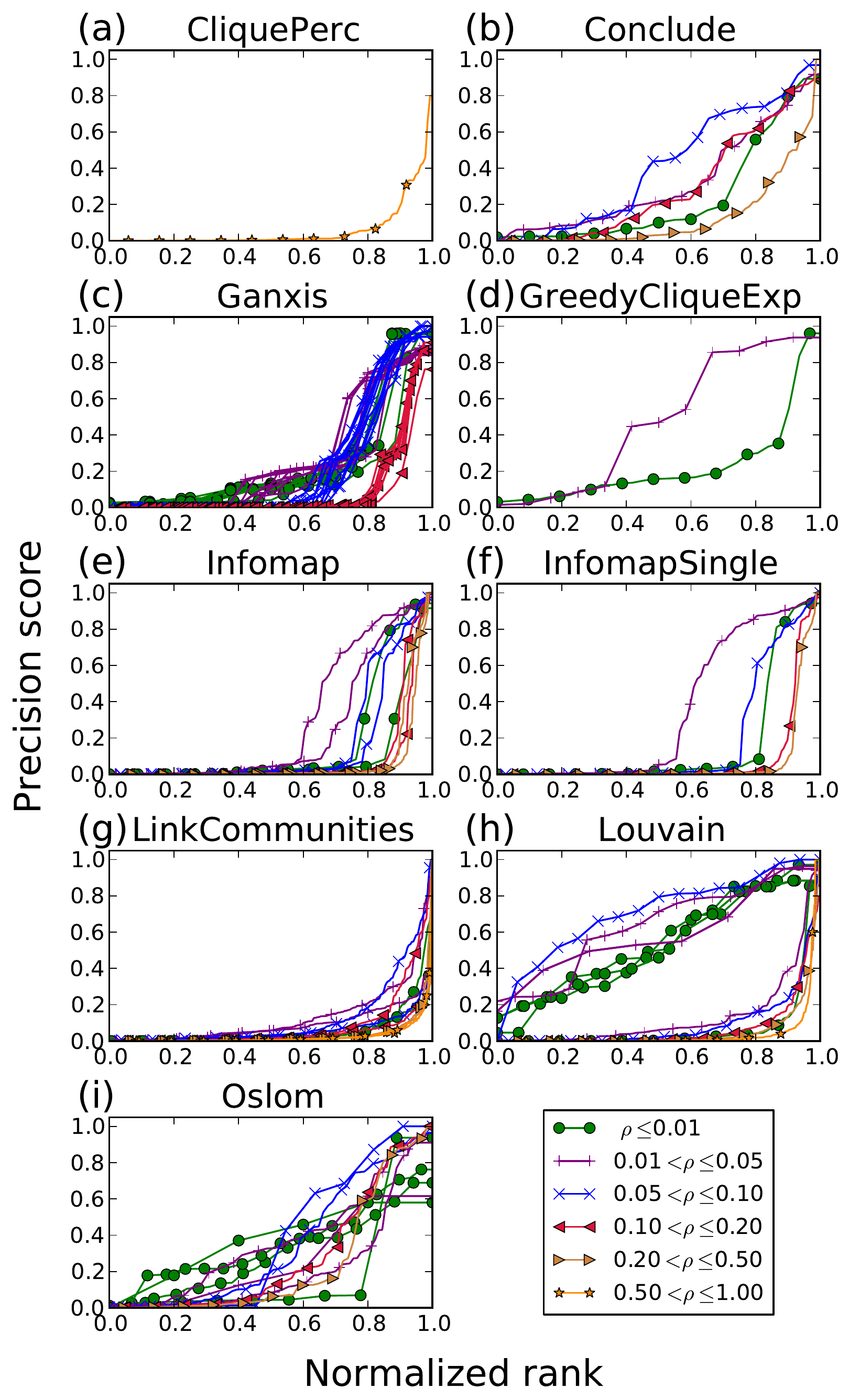}
  \caption{(Color online).  Precision of detected communities, broken
    down by density, compared to all known \DS{as-caida} groups.
    For further information, see the caption of
    Fig.~\ref{fig:as-caida-prec-size}.}
  \label{fig:as-caida-prec-den}
\end{minipage}\hfill
\begin{minipage}[t]{\columnwidth}
  \centering
  \includegraphics[width=\columnwidth]{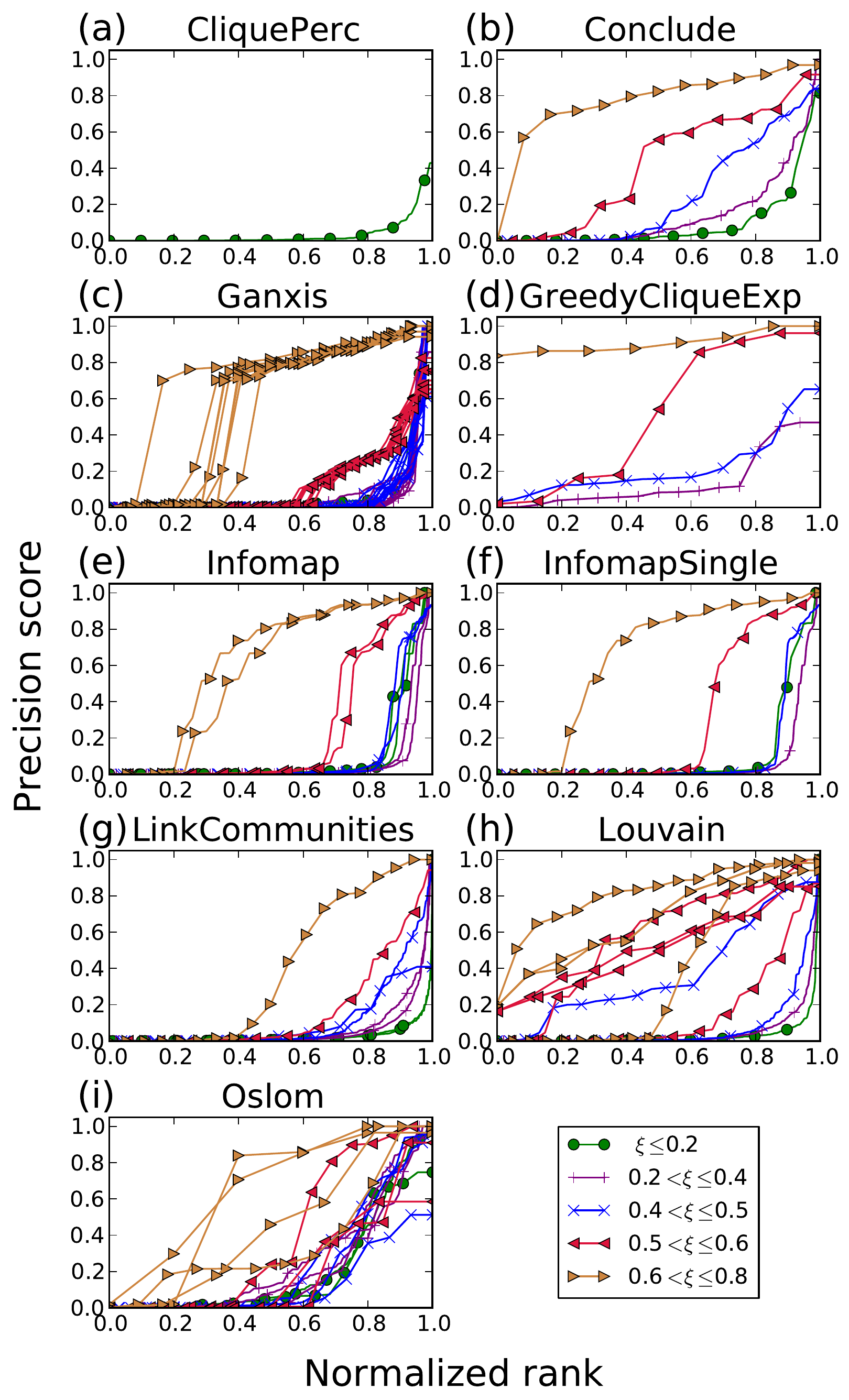}
  \caption{(Color online).  Precision of detected communities, broken
    down by community embeddedness, compared to all known
    \DS{as-caida} groups.  For further information, see the
    caption of Fig.~\ref{fig:as-caida-prec-size}. In contrast to
    Fig.~\ref{fig:as-caida-recl-ce}, no particular embeddedness
    predicts a higher performance.}
  \label{fig:as-caida-prec-ce}
\end{minipage}
\end{figure*}

\begin{figure*}
\centering
\begin{minipage}[t]{\columnwidth}
  \centering
  \includegraphics[width=\columnwidth]{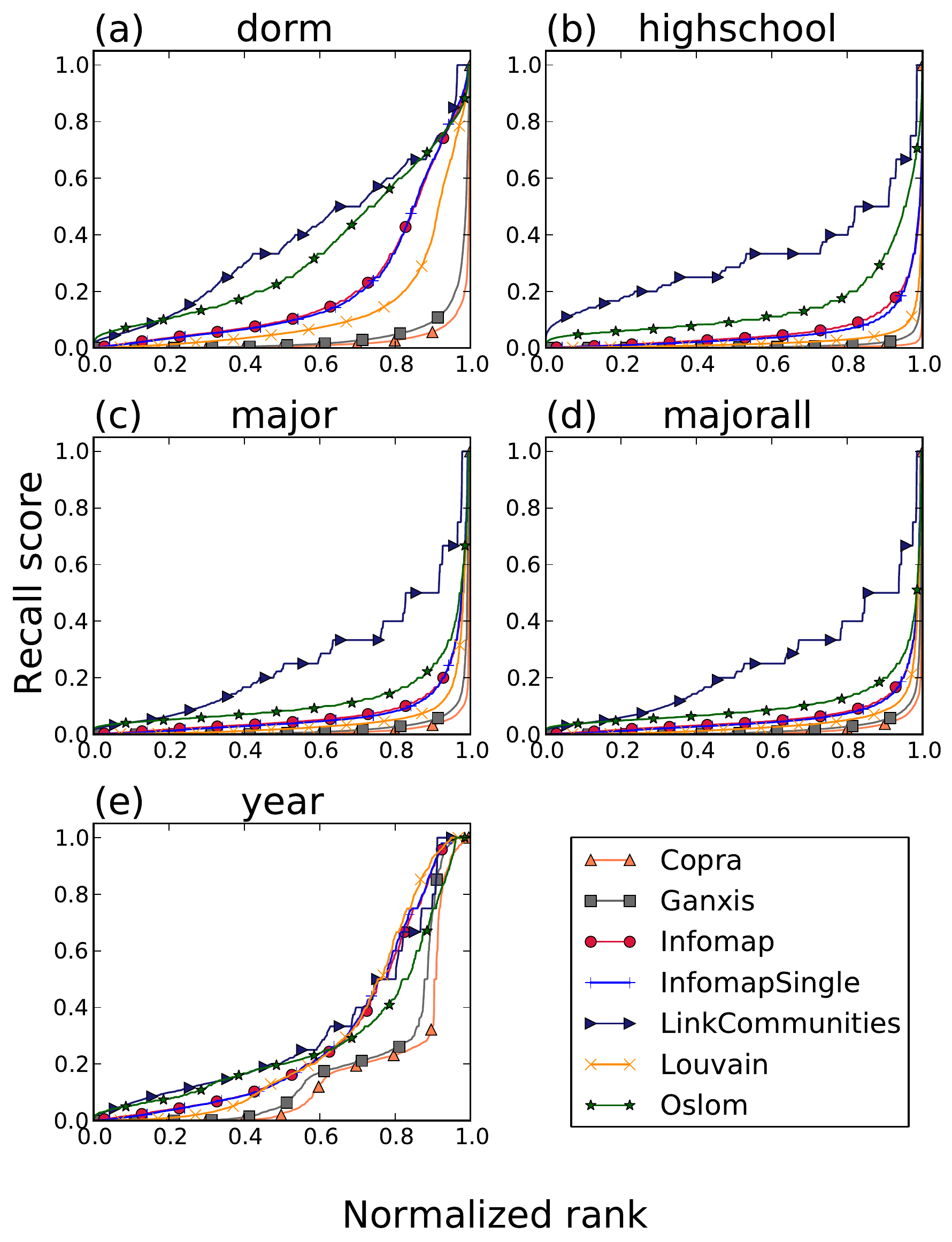}
  \caption{(Color online) Recall of metadata groups of
    \texttt{fb100}.
Each diagram refers to a grouping of the students based on one
specific feature (e.g.\ their dorm, top left).
    We see that few groups of any type of metadata are found by any
    of the community detection methods.}
  \label{fig:fb100-recl}
\end{minipage}\hfill
\begin{minipage}[t]{\columnwidth}
  \centering
  \includegraphics[width=\columnwidth]{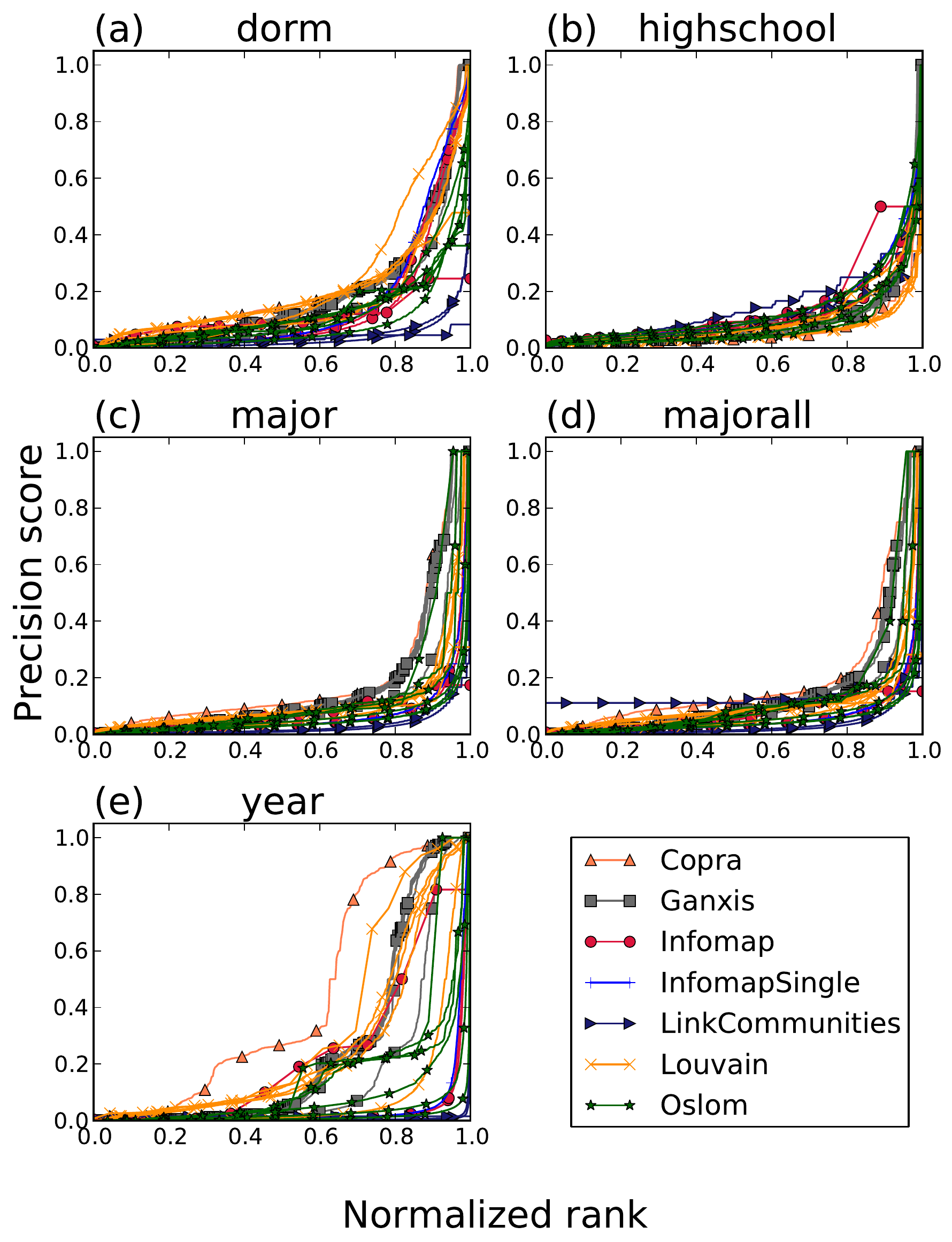}
  \caption{(Color online) Precision of each partition level of several community
    detection algorithms with respect to metadata groups of
    \texttt{fb100}, corresponding to a single feature of the students.  
    For those algorithms returning multiple detected levels,
    each level is plotted separately in order to see the
    performance of each detected layer individually.  This produces a
    field of lines, but it is sufficient to see that there
    are no outliers in performance.}
  \label{fig:fb100-prec}
\end{minipage}
\end{figure*}

We look at the properties of \textit{group size},
\textit{group density}, and \textit{group
  embeddedness} and see if any of these are indicative of a type of
group with a greater predictive power for either recall or
precision.  For a group of $n$ nodes, sum of internal degrees
\kin, sum of total degrees of \ktot, we define
the density $\rho$ as
\begin{equation}
  \rho = {\kin \over \frac{1}{2}n(n-1)}
\end{equation}
and the group embeddedness $\xi$ as
\begin{equation}
  \xi = {\kin \over \ktot}.
\end{equation}
Because some bins (parameter ranges) may have very little data, such as only one
group, we only plot bins that have at least 5 groups and
whose sum of group sizes is at least $1\%$ of the network.
Furthermore, some community detection methods return multiple covers
of the system, from
different input parameters (see Appendix \ref{sec:appendix-methods}).  When
computing recall, a known group is matched to every detected
community regardless of its detected layer or size, density, or embeddedness.  When computing precision,
one could ask if any one particular layer would have greater
predictive power than all layers taken together.  To show this, we
plot the precision of each detected layer separately.  If one
particular layer or set of parameters was very good, then we could
see one line above the rest.  As we will see, there are no significant
outliers, so the identity of each line does not matter.  This
procedure is performed on the precision plots from
Fig.~\ref{fig:as-caida-recl-size} to \ref{fig:fb100-prec}.

In Fig.~\ref{fig:as-caida-recl-size}--\ref{fig:as-caida-prec-ce},
we see the recall and precision of the \texttt{as-caida} dataset
broken down by the group properties above.  
\CDA{Copra} and \CDA{Demon} did not return
sufficient communities in each bin to perform a meaningful analysis,
so their results are not shown. In
Fig.~\ref{fig:as-caida-recl-size} and
\ref{fig:as-caida-prec-size}, we see the recall (precision)
of groups as a function of the size of the known (detected)
group.  We are able to see some variations in the performance.
Most methods seem to do a better job in detecting large groups
than small ones. A notable exception is \CDA{LinkCommunities}, which
has the highest recall for the smallest metadata groups, although
the precision for the smallest detected communities is not the
highest. In general, the curves are quite close to each other. For 
some algorithms, such as
\CDA{GreedyCliqueExp}, \CDA{InfomapSingle},
\CDA{LinkCommunities} and \CDA{OSLOM}, there is a more visible spread
of the curves.

If we consider density bins, Figs.~\ref{fig:as-caida-recl-den}  and
\ref{fig:as-caida-prec-den} show a consistent pattern as that observed
for Figs.~\ref{fig:as-caida-recl-size} and
\ref{fig:as-caida-prec-size}, as link density is correlated to
group size: small groups tend to have higher link density than
large groups.

Finally, if one considers embeddedness (Figs.~\ref{fig:as-caida-recl-ce}  and
\ref{fig:as-caida-prec-ce}), both recall and precision are highest
for the most embedded groups, i.e.\ the ones most weakly attached to
the rest of the system, and systematically decreases if embeddedness
decreases. This is expected, as most algorithms look for subgraphs
which are loosely connected to the rest of the system, and high
embeddedness means high separation.


The \DS{fb100} dataset provides us with a unique opportunity to
further understand the factors which allow high community detection
performance.  It is a collection including the Facebook social networks at 100
universities, with different types of metadata to allow us to form
groups of different types.  We can see if methods can better
detect groups of a certain type.  The metadata includes: dorm
(the student residence), high school (the school of each user before
attending university), major (the student's field of study), majorall
(the student's major(s) possibly including a second major), and year
(the student's graduation year).  In
Fig.~\ref{fig:fb100-recl} (\ref{fig:fb100-prec}) we plot the
recall (precision) of various methods with respect to groups
corresponding to each of the above attributes, averaged over the 100
universities included in this dataset.  None of 
the students' features appears to generate well recoverable
groups. \texttt{LinkCommunities}
appears to have a higher recall than most methods, for each grouping
of the students, but it has much lower precision, due to
the much bigger number of detected groups.

\begin{thebibliography}{54}
\expandafter\ifx\csname natexlab\endcsname\relax\def\natexlab#1{#1}\fi
\expandafter\ifx\csname bibnamefont\endcsname\relax
  \def\bibnamefont#1{#1}\fi
\expandafter\ifx\csname bibfnamefont\endcsname\relax
  \def\bibfnamefont#1{#1}\fi
\expandafter\ifx\csname citenamefont\endcsname\relax
  \def\citenamefont#1{#1}\fi
\expandafter\ifx\csname url\endcsname\relax
  \def\url#1{\texttt{#1}}\fi
\expandafter\ifx\csname urlprefix\endcsname\relax\def\urlprefix{URL }\fi
\providecommand{\bibinfo}[2]{#2}
\providecommand{\eprint}[2][]{\url{#2}}

\bibitem[{\citenamefont{Fortunato}(2010)}]{fortunato10}
\bibinfo{author}{\bibfnamefont{S.}~\bibnamefont{Fortunato}},
  \bibinfo{journal}{Physics Reports} \textbf{\bibinfo{volume}{486}},
  \bibinfo{pages}{75} (\bibinfo{year}{2010}).

\bibitem[{\citenamefont{Zachary}(1977)}]{zachary77}
\bibinfo{author}{\bibfnamefont{W.~W.} \bibnamefont{Zachary}},
  \bibinfo{journal}{J. Anthropol. Res.} \textbf{\bibinfo{volume}{33}},
  \bibinfo{pages}{452} (\bibinfo{year}{1977}).

\bibitem[{\citenamefont{Lusseau}(2003)}]{lusseau03}
\bibinfo{author}{\bibfnamefont{D.}~\bibnamefont{Lusseau}},
  \bibinfo{journal}{Proc. Royal Soc. London B} \textbf{\bibinfo{volume}{270}},
  \bibinfo{pages}{S186} (\bibinfo{year}{2003}).

\bibitem[{\citenamefont{Girvan and Newman}(2002)}]{girvan02}
\bibinfo{author}{\bibfnamefont{M.}~\bibnamefont{Girvan}} \bibnamefont{and}
  \bibinfo{author}{\bibfnamefont{M.~E.} \bibnamefont{Newman}},
  \bibinfo{journal}{Proc. Natl. Acad. Sci. USA} \textbf{\bibinfo{volume}{99}},
  \bibinfo{pages}{7821} (\bibinfo{year}{2002}).

\bibitem[{\citenamefont{Yang and Leskovec}(2012)}]{yang12}
\bibinfo{author}{\bibfnamefont{J.}~\bibnamefont{Yang}} \bibnamefont{and}
  \bibinfo{author}{\bibfnamefont{J.}~\bibnamefont{Leskovec}}, in
  \emph{\bibinfo{booktitle}{Proceedings of the ACM SIGKDD Workshop on Mining
  Data Semantics}} (\bibinfo{publisher}{ACM}, \bibinfo{address}{New York, NY,
  USA}, \bibinfo{year}{2012}), MDS'12, pp. \bibinfo{pages}{3:1--3:8}, ISBN
  \bibinfo{isbn}{978-1-4503-1546-3}.

\bibitem[{\citenamefont{Yang and Leskovec}(2013)}]{yang13}
\bibinfo{author}{\bibfnamefont{J.}~\bibnamefont{Yang}} \bibnamefont{and}
  \bibinfo{author}{\bibfnamefont{J.}~\bibnamefont{Leskovec}}, in
  \emph{\bibinfo{booktitle}{Proceedings of the Sixth ACM International
  Conference on Web Search and Data Mining}} (\bibinfo{publisher}{ACM},
  \bibinfo{address}{New York, NY, USA}, \bibinfo{year}{2013}), WSDM'13, pp.
  \bibinfo{pages}{587--596}, ISBN \bibinfo{isbn}{978-1-4503-1869-3}.

\bibitem[{\citenamefont{Yang and Leskovec}(2014)}]{yang14}
\bibinfo{author}{\bibfnamefont{J.}~\bibnamefont{Yang}} \bibnamefont{and}
  \bibinfo{author}{\bibfnamefont{J.}~\bibnamefont{Leskovec}},
  \bibinfo{journal}{ACM Trans. Intell. Syst. Technol.}
  \textbf{\bibinfo{volume}{5}}, \bibinfo{pages}{26:1} (\bibinfo{year}{2014}),
  ISSN \bibinfo{issn}{2157-6904}.

\bibitem[{\citenamefont{Adamic and Glance}(2005)}]{adamic05}
\bibinfo{author}{\bibfnamefont{L.~A.} \bibnamefont{Adamic}} \bibnamefont{and}
  \bibinfo{author}{\bibfnamefont{N.}~\bibnamefont{Glance}}, in
  \emph{\bibinfo{booktitle}{LinkKDD '05: Proceedings of the 3rd international
  workshop on Link discovery}} (\bibinfo{publisher}{ACM Press},
  \bibinfo{address}{New York, NY, USA}, \bibinfo{year}{2005}), pp.
  \bibinfo{pages}{36--43}.

\bibitem[{\citenamefont{Krebs}()}]{krebs06}
\bibinfo{author}{\bibfnamefont{V.}~\bibnamefont{Krebs}},
  \bibinfo{note}{unpublished from website},
  \urlprefix\url{http://www.orgnet.com/}.

\bibitem[{\citenamefont{Lancichinetti et~al.}(2008)\citenamefont{Lancichinetti,
  Fortunato, and Radicchi}}]{lancichinetti08}
\bibinfo{author}{\bibfnamefont{A.}~\bibnamefont{Lancichinetti}},
  \bibinfo{author}{\bibfnamefont{S.}~\bibnamefont{Fortunato}},
  \bibnamefont{and} \bibinfo{author}{\bibfnamefont{F.}~\bibnamefont{Radicchi}},
  \bibinfo{journal}{Phys. Rev. E} \textbf{\bibinfo{volume}{78}},
  \bibinfo{eid}{046110} (\bibinfo{year}{2008}).

\bibitem[{\citenamefont{Garfinkel}(1995)}]{garfinkel95}
\bibinfo{author}{\bibfnamefont{S.}~\bibnamefont{Garfinkel}},
  \emph{\bibinfo{title}{PGP: pretty good privacy}}
  (\bibinfo{publisher}{O'Reilly Media, Inc.}, \bibinfo{year}{1995}).

\bibitem[{cai(2013-08-01)}]{caida-as-rel}
\emph{\bibinfo{title}{The caida as relationships dataset}}
  (\bibinfo{year}{2013-08-01}),
  \urlprefix\url{http://www.caida.org/data/as-relationships/}.

\bibitem[{cai(2013-01-01 to 2013-11-25)}]{caida-ip-routed}
\emph{\bibinfo{title}{The ipv4 routed /24 as links dataset}}
  (\bibinfo{year}{2013-01-01 to 2013-11-25}),
  \urlprefix\url{http://www.caida.org/data/active//ipv4_routed_topology_aslinks_dataset.xml}.

\bibitem[{\citenamefont{Leskovec et~al.}(2007)\citenamefont{Leskovec,
      Adamic, and Huberman}}]{amazon}
\bibinfo{author}{\bibfnamefont{J.} \bibnamefont{Leskovec}},
  \bibinfo{author}{\bibfnamefont{L.~A.}~\bibnamefont{Adamic}},
  \bibnamefont{and} \bibinfo{author}{\bibfnamefont{B.~A.}~\bibnamefont{Huberman}},
  \bibinfo{journal}{ACM Transactions on the Web} \textbf{\bibinfo{volume}{1}}, \bibinfo{pages}{5}
  (\bibinfo{year}{2007}).

\bibitem[{\citenamefont{Aiello et~al.}(2012)\citenamefont{Aiello, Deplano,
  Schifanella, and Ruffo}}]{aiello12}
\bibinfo{author}{\bibfnamefont{L.~M.} \bibnamefont{Aiello}},
  \bibinfo{author}{\bibfnamefont{M.}~\bibnamefont{Deplano}},
  \bibinfo{author}{\bibfnamefont{R.}~\bibnamefont{Schifanella}},
  \bibnamefont{and}
  \bibinfo{author}{\bibfnamefont{G.}~\bibnamefont{Ruffo}}, , in
  \emph{\bibinfo{booktitle}{Sixth International AAAI Conference on
      Weblogs and Social Media}} (\bibinfo{organization}{AAAI},
  \bibinfo{year}{2012}), pp.  \bibinfo{pages}{10--17}.

\bibitem[{\citenamefont{Aiello et~al.}(2010)\citenamefont{Aiello, Barrat,
  Cattuto, Ruffo, and Schifanella}}]{aiello10}
\bibinfo{author}{\bibfnamefont{L.~M.} \bibnamefont{Aiello}},
  \bibinfo{author}{\bibfnamefont{A.}~\bibnamefont{Barrat}},
  \bibinfo{author}{\bibfnamefont{C.}~\bibnamefont{Cattuto}},
  \bibinfo{author}{\bibfnamefont{G.}~\bibnamefont{Ruffo}}, \bibnamefont{and}
  \bibinfo{author}{\bibfnamefont{R.}~\bibnamefont{Schifanella}}, in
  \emph{\bibinfo{booktitle}{Social Computing (SocialCom), 2010 IEEE Second
  International Conference on}} (\bibinfo{organization}{IEEE},
  \bibinfo{year}{2010}), pp. \bibinfo{pages}{249--256}.

\bibitem[{\citenamefont{Backstrom et~al.}(2006)\citenamefont{Backstrom,
  Huttenlocher, Kleinberg, and Lan}}]{backstrom06}
\bibinfo{author}{\bibfnamefont{L.}~\bibnamefont{Backstrom}},
  \bibinfo{author}{\bibfnamefont{D.}~\bibnamefont{Huttenlocher}},
  \bibinfo{author}{\bibfnamefont{J.}~\bibnamefont{Kleinberg}},
  \bibnamefont{and} \bibinfo{author}{\bibfnamefont{X.}~\bibnamefont{Lan}}, in
  \emph{\bibinfo{booktitle}{KDD '06: Proceedings of the 12th ACM SIGKDD
  international conference on Knowledge discovery and data mining}}
  (\bibinfo{publisher}{ACM}, \bibinfo{address}{New York, NY, USA},
  \bibinfo{year}{2006}), pp. \bibinfo{pages}{44--54}.

\bibitem[{\citenamefont{Traud et~al.}(2012)\citenamefont{Traud, Mucha, and
  Porter}}]{traud12}
\bibinfo{author}{\bibfnamefont{A.~L.} \bibnamefont{Traud}},
  \bibinfo{author}{\bibfnamefont{P.~J.} \bibnamefont{Mucha}}, \bibnamefont{and}
  \bibinfo{author}{\bibfnamefont{M.~A.} \bibnamefont{Porter}},
  \bibinfo{journal}{Physica A: Statistical Mechanics and its Applications}
  \textbf{\bibinfo{volume}{391}}, \bibinfo{pages}{4165 }
  (\bibinfo{year}{2012}).

\bibitem[{\citenamefont{Mislove et~al.}(2007)\citenamefont{Mislove, Marcon,
  Gummadi, Druschel, and Bhattacharjee}}]{mislove07}
\bibinfo{author}{\bibfnamefont{A.}~\bibnamefont{Mislove}},
  \bibinfo{author}{\bibfnamefont{M.}~\bibnamefont{Marcon}},
  \bibinfo{author}{\bibfnamefont{K.~P.} \bibnamefont{Gummadi}},
  \bibinfo{author}{\bibfnamefont{P.}~\bibnamefont{Druschel}}, \bibnamefont{and}
  \bibinfo{author}{\bibfnamefont{B.}~\bibnamefont{Bhattacharjee}}, in
  \emph{\bibinfo{booktitle}{Proceedings of the 7th ACM SIGCOMM conference on
  Internet measurement}} (\bibinfo{organization}{ACM}, \bibinfo{year}{2007}),
  pp. \bibinfo{pages}{29--42}.

\bibitem[{\citenamefont{Blondel et~al.}(2008)\citenamefont{Blondel, Guillaume,
  Lambiotte, and Lefebvre}}]{blondel08}
\bibinfo{author}{\bibfnamefont{V.~D.} \bibnamefont{Blondel}},
  \bibinfo{author}{\bibfnamefont{J.-L.} \bibnamefont{Guillaume}},
  \bibinfo{author}{\bibfnamefont{R.}~\bibnamefont{Lambiotte}},
  \bibnamefont{and} \bibinfo{author}{\bibfnamefont{E.}~\bibnamefont{Lefebvre}},
  \bibinfo{journal}{J. Stat. Mech.} \textbf{\bibinfo{volume}{P10008}}
  (\bibinfo{year}{2008}).

\bibitem[{\citenamefont{Rosvall and Bergstrom}(2011)}]{rosvall11}
\bibinfo{author}{\bibfnamefont{M.}~\bibnamefont{Rosvall}} \bibnamefont{and}
  \bibinfo{author}{\bibfnamefont{C.~T.} \bibnamefont{Bergstrom}},
  \bibinfo{journal}{PLoS ONE} \textbf{\bibinfo{volume}{6}},
  \bibinfo{pages}{e18209} (\bibinfo{year}{2011}).

\bibitem[{\citenamefont{{Rosvall} and {Bergstrom}}(2008)}]{rosvall08}
\bibinfo{author}{\bibfnamefont{M.}~\bibnamefont{{Rosvall}}} \bibnamefont{and}
  \bibinfo{author}{\bibfnamefont{C.~T.} \bibnamefont{{Bergstrom}}},
  \bibinfo{journal}{Proc. Natl. Acad. Sci. USA} \textbf{\bibinfo{volume}{105}},
  \bibinfo{pages}{1118} (\bibinfo{year}{2008}).

\bibitem[{\citenamefont{Ahn et~al.}(2010)\citenamefont{Ahn, Bagrow, and
  Lehmann}}]{ahn10}
\bibinfo{author}{\bibfnamefont{Y.-Y.} \bibnamefont{Ahn}},
  \bibinfo{author}{\bibfnamefont{J.~P.} \bibnamefont{Bagrow}},
  \bibnamefont{and} \bibinfo{author}{\bibfnamefont{S.}~\bibnamefont{Lehmann}},
  \bibinfo{journal}{Nature} \textbf{\bibinfo{volume}{466}},
  \bibinfo{pages}{761} (\bibinfo{year}{2010}).

\bibitem[{\citenamefont{{Palla} et~al.}(2005)\citenamefont{{Palla},
  {Der{\'e}nyi}, {Farkas}, and {Vicsek}}}]{palla05}
\bibinfo{author}{\bibfnamefont{G.}~\bibnamefont{{Palla}}},
  \bibinfo{author}{\bibfnamefont{I.}~\bibnamefont{{Der{\'e}nyi}}},
  \bibinfo{author}{\bibfnamefont{I.}~\bibnamefont{{Farkas}}}, \bibnamefont{and}
  \bibinfo{author}{\bibfnamefont{T.}~\bibnamefont{{Vicsek}}},
  \bibinfo{journal}{Nature} \textbf{\bibinfo{volume}{435}},
  \bibinfo{pages}{814} (\bibinfo{year}{2005}).

\bibitem[{\citenamefont{Reid et~al.}(2012)\citenamefont{Reid, McDaid, and
  Hurley}}]{reid12}
\bibinfo{author}{\bibfnamefont{F.}~\bibnamefont{Reid}},
  \bibinfo{author}{\bibfnamefont{A.}~\bibnamefont{McDaid}}, \bibnamefont{and}
  \bibinfo{author}{\bibfnamefont{N.}~\bibnamefont{Hurley}}, in
  \emph{\bibinfo{booktitle}{Advances in Social Networks Analysis and Mining
  (ASONAM), 2012 IEEE/ACM International Conference on}}
  (\bibinfo{organization}{IEEE}, \bibinfo{year}{2012}), pp.
  \bibinfo{pages}{274--281}.

\bibitem[{\citenamefont{De~Meo et~al.}(2014)\citenamefont{De~Meo, Ferrara,
  Fiumara, and Provetti}}]{demeo14}
\bibinfo{author}{\bibfnamefont{P.}~\bibnamefont{De~Meo}},
  \bibinfo{author}{\bibfnamefont{E.}~\bibnamefont{Ferrara}},
  \bibinfo{author}{\bibfnamefont{G.}~\bibnamefont{Fiumara}}, \bibnamefont{and}
  \bibinfo{author}{\bibfnamefont{A.}~\bibnamefont{Provetti}},
  \bibinfo{journal}{Journal of Computer and System Sciences}
  \textbf{\bibinfo{volume}{80}}, \bibinfo{pages}{72} (\bibinfo{year}{2014}).

\bibitem[{\citenamefont{Gregory}(2010)}]{gregory10}
\bibinfo{author}{\bibfnamefont{S.}~\bibnamefont{Gregory}},
  \bibinfo{journal}{New J. Phys.} \textbf{\bibinfo{volume}{12}},
  \bibinfo{pages}{103018} (\bibinfo{year}{2010}).

\bibitem[{\citenamefont{Coscia et~al.}(2012)\citenamefont{Coscia, Rossetti,
  Giannotti, and Pedreschi}}]{coscia12}
\bibinfo{author}{\bibfnamefont{M.}~\bibnamefont{Coscia}},
  \bibinfo{author}{\bibfnamefont{G.}~\bibnamefont{Rossetti}},
  \bibinfo{author}{\bibfnamefont{F.}~\bibnamefont{Giannotti}},
  \bibnamefont{and}
  \bibinfo{author}{\bibfnamefont{D.}~\bibnamefont{Pedreschi}}, in
  \emph{\bibinfo{booktitle}{Proceedings of the 18th ACM SIGKDD International
  Conference on Knowledge Discovery and Data Mining}}
  (\bibinfo{publisher}{ACM}, \bibinfo{address}{New York, NY, USA},
  \bibinfo{year}{2012}), KDD'12, pp. \bibinfo{pages}{615--623}.

\bibitem[{\citenamefont{Xie and Szymanski}(2012)}]{xie12}
\bibinfo{author}{\bibfnamefont{J.}~\bibnamefont{Xie}} \bibnamefont{and}
  \bibinfo{author}{\bibfnamefont{B.~K.} \bibnamefont{Szymanski}}, in
  \emph{\bibinfo{booktitle}{Advances in Knowledge Discovery and Data Mining}}
  (\bibinfo{publisher}{Springer}, \bibinfo{year}{2012}), pp.
  \bibinfo{pages}{25--36}.

\bibitem[{\citenamefont{Lee et~al.}(2010)\citenamefont{Lee, Reid, McDaid, and
  Hurley}}]{lee10}
\bibinfo{author}{\bibfnamefont{C.}~\bibnamefont{Lee}},
  \bibinfo{author}{\bibfnamefont{F.}~\bibnamefont{Reid}},
  \bibinfo{author}{\bibfnamefont{A.}~\bibnamefont{McDaid}}, \bibnamefont{and}
  \bibinfo{author}{\bibfnamefont{N.}~\bibnamefont{Hurley}},
  \bibinfo{journal}{arXiv preprint arXiv:1002.1827}  (\bibinfo{year}{2010}).

\bibitem[{\citenamefont{{Danon} et~al.}(2005)\citenamefont{{Danon},
  {D{\'{\i}}az-Guilera}, {Duch}, and {Arenas}}}]{danon05}
\bibinfo{author}{\bibfnamefont{L.}~\bibnamefont{{Danon}}},
  \bibinfo{author}{\bibfnamefont{A.}~\bibnamefont{{D{\'{\i}}az-Guilera}}},
  \bibinfo{author}{\bibfnamefont{J.}~\bibnamefont{{Duch}}}, \bibnamefont{and}
  \bibinfo{author}{\bibfnamefont{A.}~\bibnamefont{{Arenas}}},
  \bibinfo{journal}{J. Stat. Mech.} \textbf{\bibinfo{volume}{P09008}}
  (\bibinfo{year}{2005}).

\bibitem[{\citenamefont{Lancichinetti et~al.}(2009)\citenamefont{Lancichinetti,
  Fortunato, and Kertesz}}]{lancichinetti09}
\bibinfo{author}{\bibfnamefont{A.}~\bibnamefont{Lancichinetti}},
  \bibinfo{author}{\bibfnamefont{S.}~\bibnamefont{Fortunato}},
  \bibnamefont{and} \bibinfo{author}{\bibfnamefont{J.}~\bibnamefont{Kertesz}},
  \bibinfo{journal}{New J. Phys.} \textbf{\bibinfo{volume}{11}},
  \bibinfo{pages}{033015} (\bibinfo{year}{2009}).

\bibitem[{\citenamefont{Abrahao et~al.}(2012)\citenamefont{Abrahao,
  Soundarajan, Hopcroft, and Kleinberg}}]{abrahao12}
\bibinfo{author}{\bibfnamefont{B.}~\bibnamefont{Abrahao}},
  \bibinfo{author}{\bibfnamefont{S.}~\bibnamefont{Soundarajan}},
  \bibinfo{author}{\bibfnamefont{J.}~\bibnamefont{Hopcroft}}, \bibnamefont{and}
  \bibinfo{author}{\bibfnamefont{R.}~\bibnamefont{Kleinberg}}, in
  \emph{\bibinfo{booktitle}{Proceedings of the 18th ACM SIGKDD International
  Conference on Knowledge Discovery and Data Mining}}
  (\bibinfo{publisher}{ACM}, \bibinfo{address}{New York, NY, USA},
  \bibinfo{year}{2012}), KDD '12, pp. \bibinfo{pages}{624--632}, ISBN
  \bibinfo{isbn}{978-1-4503-1462-6}.

\bibitem[{\citenamefont{Ester et~al.}(2006)\citenamefont{Ester, Ge, Gao, Hu,
  and Ben-Moshe}}]{ester06}
\bibinfo{author}{\bibfnamefont{M.}~\bibnamefont{Ester}},
  \bibinfo{author}{\bibfnamefont{R.}~\bibnamefont{Ge}},
  \bibinfo{author}{\bibfnamefont{B.~J.} \bibnamefont{Gao}},
  \bibinfo{author}{\bibfnamefont{Z.}~\bibnamefont{Hu}}, \bibnamefont{and}
  \bibinfo{author}{\bibfnamefont{B.}~\bibnamefont{Ben-Moshe}}, in
  \emph{\bibinfo{booktitle}{SDM'06}}, edited by
  \bibinfo{editor}{\bibfnamefont{J.}~\bibnamefont{Ghosh}},
  \bibinfo{editor}{\bibfnamefont{D.}~\bibnamefont{Lambert}},
  \bibinfo{editor}{\bibfnamefont{D.~B.} \bibnamefont{Skillicorn}},
  \bibnamefont{and}
  \bibinfo{editor}{\bibfnamefont{J.}~\bibnamefont{Srivastava}}
  (\bibinfo{publisher}{SIAM}, \bibinfo{year}{2006}).

\bibitem[{\citenamefont{Liu et~al.}(2009)\citenamefont{Liu, Niculescu-Mizil,
  and Gryc}}]{liu09}
\bibinfo{author}{\bibfnamefont{Y.}~\bibnamefont{Liu}},
  \bibinfo{author}{\bibfnamefont{A.}~\bibnamefont{Niculescu-Mizil}},
  \bibnamefont{and} \bibinfo{author}{\bibfnamefont{W.}~\bibnamefont{Gryc}}, in
  \emph{\bibinfo{booktitle}{Proceedings of the 26th Annual International
  Conference on Machine Learning}} (\bibinfo{publisher}{ACM},
  \bibinfo{address}{New York, NY, USA}, \bibinfo{year}{2009}), ICML '09, pp.
  \bibinfo{pages}{665--672}, ISBN \bibinfo{isbn}{978-1-60558-516-1}.

\bibitem[{\citenamefont{Moser et~al.}(2009)\citenamefont{Moser, Colak, Rafiey,
  and Ester}}]{moser09}
\bibinfo{author}{\bibfnamefont{F.}~\bibnamefont{Moser}},
  \bibinfo{author}{\bibfnamefont{R.}~\bibnamefont{Colak}},
  \bibinfo{author}{\bibfnamefont{A.}~\bibnamefont{Rafiey}}, \bibnamefont{and}
  \bibinfo{author}{\bibfnamefont{M.}~\bibnamefont{Ester}}, in
  \emph{\bibinfo{booktitle}{SDM'09}} (\bibinfo{publisher}{SIAM},
  \bibinfo{year}{2009}), pp. \bibinfo{pages}{593--604}.

\bibitem[{\citenamefont{Zhou et~al.}(2009)\citenamefont{Zhou, Cheng, and
  Yu}}]{zhou09}
\bibinfo{author}{\bibfnamefont{Y.}~\bibnamefont{Zhou}},
  \bibinfo{author}{\bibfnamefont{H.}~\bibnamefont{Cheng}}, \bibnamefont{and}
  \bibinfo{author}{\bibfnamefont{J.~X.} \bibnamefont{Yu}},
  \bibinfo{journal}{Proc. VLDB Endow.} \textbf{\bibinfo{volume}{2}},
  \bibinfo{pages}{718} (\bibinfo{year}{2009}), ISSN \bibinfo{issn}{2150-8097}.

\bibitem[{\citenamefont{Tang et~al.}(2009)\citenamefont{Tang, Wang, and
  Liu}}]{tang09}
\bibinfo{author}{\bibfnamefont{L.}~\bibnamefont{Tang}},
  \bibinfo{author}{\bibfnamefont{X.}~\bibnamefont{Wang}}, \bibnamefont{and}
  \bibinfo{author}{\bibfnamefont{H.}~\bibnamefont{Liu}}, in
  \emph{\bibinfo{booktitle}{IEEE International Conference on Data Mining
  (ICDM'09)}} (\bibinfo{year}{2009}), pp. \bibinfo{pages}{503--512}.

\bibitem[{\citenamefont{Silva et~al.}(2010)\citenamefont{Silva, Meira, and
  Zaki}}]{silva10}
\bibinfo{author}{\bibfnamefont{A.}~\bibnamefont{Silva}},
  \bibinfo{author}{\bibfnamefont{W.}~\bibnamefont{Meira}, \bibfnamefont{Jr.}},
  \bibnamefont{and} \bibinfo{author}{\bibfnamefont{M.~J.} \bibnamefont{Zaki}},
  in \emph{\bibinfo{booktitle}{Proceedings of the Eighth Workshop on Mining and
  Learning with Graphs}} (\bibinfo{publisher}{ACM}, \bibinfo{address}{New York,
  NY, USA}, \bibinfo{year}{2010}), MLG '10, pp. \bibinfo{pages}{119--126}.

\bibitem[{\citenamefont{Balasubramanyan and Cohen}(2011)}]{balasubramanyan11}
\bibinfo{author}{\bibfnamefont{R.}~\bibnamefont{Balasubramanyan}}
  \bibnamefont{and} \bibinfo{author}{\bibfnamefont{W.~W.} \bibnamefont{Cohen}},
  in \emph{\bibinfo{booktitle}{SDM'11}} (\bibinfo{publisher}{SIAM / Omnipress},
  \bibinfo{year}{2011}), pp. \bibinfo{pages}{450--461}.

\bibitem[{\citenamefont{Atzmueller and Mitzlaff}(2011)}]{atzmueller11}
\bibinfo{author}{\bibfnamefont{M.}~\bibnamefont{Atzmueller}} \bibnamefont{and}
  \bibinfo{author}{\bibfnamefont{F.}~\bibnamefont{Mitzlaff}}, in
  \emph{\bibinfo{booktitle}{Proceedings of the Twenty-Fourth International
  Florida Artificial Intelligence Research Society Conference, May 18-20, 2011,
  Palm Beach, Florida, USA}}, edited by \bibinfo{editor}{\bibfnamefont{R.~C.}
  \bibnamefont{Murray}} \bibnamefont{and} \bibinfo{editor}{\bibfnamefont{P.~M.}
  \bibnamefont{McCarthy}} (\bibinfo{publisher}{AAAI Press},
  \bibinfo{year}{2011}), pp. \bibinfo{pages}{459--464}.

\bibitem[{\citenamefont{Akoglu et~al.}(2012)\citenamefont{Akoglu, Tong, Meeder,
  and Faloutsos}}]{akoglu12}
\bibinfo{author}{\bibfnamefont{L.}~\bibnamefont{Akoglu}},
  \bibinfo{author}{\bibfnamefont{H.}~\bibnamefont{Tong}},
  \bibinfo{author}{\bibfnamefont{B.}~\bibnamefont{Meeder}}, \bibnamefont{and}
  \bibinfo{author}{\bibfnamefont{C.}~\bibnamefont{Faloutsos}}, in
  \emph{\bibinfo{booktitle}{SDM'12}} (\bibinfo{publisher}{SIAM / Omnipress},
  \bibinfo{year}{2012}), pp. \bibinfo{pages}{439--450}.

\bibitem[{\citenamefont{Bonchi et~al.}(2012)\citenamefont{Bonchi, Gionis,
  Gullo, and Ukkonen}}]{bonchi12}
\bibinfo{author}{\bibfnamefont{F.}~\bibnamefont{Bonchi}},
  \bibinfo{author}{\bibfnamefont{A.}~\bibnamefont{Gionis}},
  \bibinfo{author}{\bibfnamefont{F.}~\bibnamefont{Gullo}}, \bibnamefont{and}
  \bibinfo{author}{\bibfnamefont{A.}~\bibnamefont{Ukkonen}}, in
  \emph{\bibinfo{booktitle}{Proceedings of the 18th ACM SIGKDD International
  Conference on Knowledge Discovery and Data Mining}}
  (\bibinfo{publisher}{ACM}, \bibinfo{address}{New York, NY, USA},
  \bibinfo{year}{2012}), KDD '12, pp. \bibinfo{pages}{1321--1329}.

\bibitem[{\citenamefont{Sun et~al.}(2012)\citenamefont{Sun, Aggarwal, and
  Han}}]{sun12}
\bibinfo{author}{\bibfnamefont{Y.}~\bibnamefont{Sun}},
  \bibinfo{author}{\bibfnamefont{C.~C.} \bibnamefont{Aggarwal}},
  \bibnamefont{and} \bibinfo{author}{\bibfnamefont{J.}~\bibnamefont{Han}},
  \bibinfo{journal}{Proc. VLDB Endow.} \textbf{\bibinfo{volume}{5}},
  \bibinfo{pages}{394} (\bibinfo{year}{2012}).

\bibitem[{\citenamefont{Xu et~al.}(2012)\citenamefont{Xu, Ke, Wang, Cheng, and
  Cheng}}]{xu12}
\bibinfo{author}{\bibfnamefont{Z.}~\bibnamefont{Xu}},
  \bibinfo{author}{\bibfnamefont{Y.}~\bibnamefont{Ke}},
  \bibinfo{author}{\bibfnamefont{Y.}~\bibnamefont{Wang}},
  \bibinfo{author}{\bibfnamefont{H.}~\bibnamefont{Cheng}}, \bibnamefont{and}
  \bibinfo{author}{\bibfnamefont{J.}~\bibnamefont{Cheng}}, in
  \emph{\bibinfo{booktitle}{Proceedings of the 2012 ACM SIGMOD International
  Conference on Management of Data}} (\bibinfo{publisher}{ACM},
  \bibinfo{address}{New York, NY, USA}, \bibinfo{year}{2012}), SIGMOD '12, pp.
  \bibinfo{pages}{505--516}.

\bibitem[{\citenamefont{Barbieri et~al.}(2013)\citenamefont{Barbieri, Bonchi,
  and Manco}}]{barbieri13}
\bibinfo{author}{\bibfnamefont{N.}~\bibnamefont{Barbieri}},
  \bibinfo{author}{\bibfnamefont{F.}~\bibnamefont{Bonchi}}, \bibnamefont{and}
  \bibinfo{author}{\bibfnamefont{G.}~\bibnamefont{Manco}}, in
  \emph{\bibinfo{booktitle}{Proceedings of the Sixth ACM International
  Conference on Web Search and Data Mining}} (\bibinfo{publisher}{ACM},
  \bibinfo{address}{New York, NY, USA}, \bibinfo{year}{2013}), WSDM'13, pp.
  \bibinfo{pages}{33--42}.

\bibitem[{\citenamefont{Ruan et~al.}(2013)\citenamefont{Ruan, Fuhry, and
  Parthasarathy}}]{ruan13}
\bibinfo{author}{\bibfnamefont{Y.}~\bibnamefont{Ruan}},
  \bibinfo{author}{\bibfnamefont{D.}~\bibnamefont{Fuhry}}, \bibnamefont{and}
  \bibinfo{author}{\bibfnamefont{S.}~\bibnamefont{Parthasarathy}}, in
  \emph{\bibinfo{booktitle}{Proceedings of the 22Nd International Conference on
  World Wide Web}} (\bibinfo{publisher}{International World Wide Web
  Conferences Steering Committee}, \bibinfo{address}{Republic and Canton of
  Geneva, Switzerland}, \bibinfo{year}{2013}), WWW '13, pp.
  \bibinfo{pages}{1089--1098}.

\bibitem[{\citenamefont{Yang et~al.}(2013)\citenamefont{Yang, Macauley, and
  Leskovec}}]{yang13b}
\bibinfo{author}{\bibfnamefont{J.}~\bibnamefont{Yang}},
  \bibinfo{author}{\bibfnamefont{J.~J.} \bibnamefont{Macauley}},
  \bibnamefont{and} \bibinfo{author}{\bibfnamefont{J.}~\bibnamefont{Leskovec}}
  (\bibinfo{year}{2013}), \eprint{arXiv:1401.7267}.

\bibitem[{\citenamefont{Pool et~al.}(2014)\citenamefont{Pool, Bonchi, and van
  Leeuwen}}]{pool14}
\bibinfo{author}{\bibfnamefont{S.}~\bibnamefont{Pool}},
  \bibinfo{author}{\bibfnamefont{F.}~\bibnamefont{Bonchi}}, \bibnamefont{and}
  \bibinfo{author}{\bibfnamefont{M.}~\bibnamefont{van Leeuwen}},
  \bibinfo{journal}{ACM Trans. Intell. Syst. Technol.}
  \textbf{\bibinfo{volume}{5}}, \bibinfo{pages}{28:1} (\bibinfo{year}{2014}).

\bibitem[{\citenamefont{Evans}(2012)}]{Evans}
\bibinfo{author}{\bibfnamefont{T.}~\bibnamefont{Evans}},
  \emph{\bibinfo{title}{American college football network files. figshare.}}
  (\bibinfo{year}{2012}),
  \urlprefix\url{http://dx.doi.org/10.6084/m9.figshare.93179}.

\bibitem[{deb()}]{debian7.1}
\emph{\bibinfo{title}{Debian gnu/linux 7.1}},
  \urlprefix\url{http://debian.org/}.

\bibitem[{\citenamefont{Barth et~al.}(1997-2013)\citenamefont{Barth, Di~Carlo,
  Hertzog, Nussbaum, Schwarz, and Jackson}}]{debianDevRef}
\bibinfo{author}{\bibfnamefont{A.}~\bibnamefont{Barth}},
  \bibinfo{author}{\bibfnamefont{A.}~\bibnamefont{Di~Carlo}},
  \bibinfo{author}{\bibfnamefont{R.}~\bibnamefont{Hertzog}},
  \bibinfo{author}{\bibfnamefont{L.}~\bibnamefont{Nussbaum}},
  \bibinfo{author}{\bibfnamefont{C.}~\bibnamefont{Schwarz}}, \bibnamefont{and}
  \bibinfo{author}{\bibfnamefont{I.}~\bibnamefont{Jackson}},
  \emph{\bibinfo{title}{Debian developer's reference}}
  (\bibinfo{year}{1997-2013}), \bibinfo{note}{v3.4.11},
  \urlprefix\url{https://www.debian.org/doc/manuals/developers-reference/}.

\bibitem[{\citenamefont{Zini}(2005)}]{zini05}
\bibinfo{author}{\bibfnamefont{E.}~\bibnamefont{Zini}}, in
  \emph{\bibinfo{booktitle}{Proceedings of the 5th annual Debian Conference}}
  (\bibinfo{year}{2005}), pp. \bibinfo{pages}{59--74},
  \urlprefix\url{http://debtags.alioth.debian.org/paper-debtags.html}.

\bibitem[{\citenamefont{Hawkinson and Bates}(1996)}]{rfc1930}
\bibinfo{author}{\bibfnamefont{J.}~\bibnamefont{Hawkinson}} \bibnamefont{and}
  \bibinfo{author}{\bibfnamefont{T.}~\bibnamefont{Bates}},
  \emph{\bibinfo{title}{Guidelines for creation, selection, and registration of
  an autonomous system (as)}} (\bibinfo{year}{1996}),
  \urlprefix\url{http://tools.ietf.org/html/rfc1930}.

\bibitem[{\citenamefont{Lancichinetti et~al.}(2011)\citenamefont{Lancichinetti,
  Radicchi, Ramasco, and Fortunato}}]{lancichinetti11}
\bibinfo{author}{\bibfnamefont{A.}~\bibnamefont{Lancichinetti}},
  \bibinfo{author}{\bibfnamefont{F.}~\bibnamefont{Radicchi}},
  \bibinfo{author}{\bibfnamefont{J.~J.} \bibnamefont{Ramasco}},
  \bibnamefont{and}
  \bibinfo{author}{\bibfnamefont{S.}~\bibnamefont{Fortunato}},
  \bibinfo{journal}{PLoS ONE} \textbf{\bibinfo{volume}{6}},
  \bibinfo{pages}{e18961} (\bibinfo{year}{2011}).

\end{thebibliography}
\end{document}